\documentclass[ journal]{vgtc}                
\usepackage{gensymb}

\usepackage[pdftex]{graphicx}
\graphicspath{{../pic/}{../jpeg/}}
\DeclareGraphicsExtensions{.pdf,.jpeg,.png}
  
\usepackage{xspace}
\usepackage[switch]{lineno}
\usepackage{enumitem}
\usepackage{rotating}
\usepackage{float,dblfloatfix}
\usepackage{wrapfig}
\setlist[enumerate]{itemsep=-3.5pt,topsep=2pt}

\usepackage{mathptmx, amsmath, amssymb}
\usepackage{times}
\usepackage{comment}
\usepackage{cite}  
\usepackage{color,colortbl,xcolor}
\usepackage[normalem]{ulem}
 \usepackage{array}

\usepackage[framemethod=TikZ]{mdframed}
\usepackage{framed}
\usepackage{tcolorbox} 

\DeclareMathSizes{1}{1}{1}{1}

\definecolor{pink}{rgb}{.98,.52,.40}
\definecolor{orange}{RGB}{227, 143, 18}
\definecolor{red}{RGB}{240, 64, 64}
\definecolor{olive}{rgb}{.25,.52,.15}
\definecolor{blue}{HTML}{2d68c4}
\definecolor{lavendar}{RGB}{115, 79, 150}
\definecolor{turquoise}{RGB}{64, 224, 208}
\definecolor{lightblue}{RGB}{238, 240, 250}
\definecolor{lightred}{RGB}{255, 241, 207}
\definecolor{lightgray}{HTML}{bbbbbb}
\definecolor{verylightgray}{RGB}{240, 240, 240}
\definecolor{gray}{RGB}{90,90,90}
\definecolor{black}{RGB}{0, 0, 0}
\definecolor{very_light_red}{HTML}{f7ebeb}
\definecolor{very_light_purple}{HTML}{edebf7}
\definecolor{hl_color}{HTML}{fff9eb}
\definecolor{hl_color2}{HTML}{e9f2ff}

\definecolor{bias_color}{HTML}{c4b84d}
\definecolor{precision_color}{HTML}{33a3a3}
\definecolor{target_color}{HTML}{c43f3f}
\definecolor{response_color}{HTML}{777a91}
\definecolor{position_color}{RGB}{102,138,196} 
\definecolor{difference_color}{RGB}{247,194,129}
\definecolor{saturation_color}{RGB}{177,177,177} 
\definecolor{area_color}{RGB}{147,139,193}
\definecolor{length_color}{RGB}{146,183,121}
\definecolor{orientation_color}{RGB}{107,189,214}
                   
\definecolor{cmcolor}{RGB}{200, 100, 0}
\definecolor{sfcolor}{RGB}{179, 64, 64}
\definecolor{fycolor}{RGB}{131, 121, 181}
\definecolor{tbcolor}{RGB}{79,174,205}
\definecolor{bgcolor}{RGB}{230, 230, 230}

\definecolor{model_responses_color}{RGB}{209, 107, 136}
\definecolor{model_nu_color}{RGB}{74, 163, 74}
\definecolor{model_normal_color}{RGB}{91, 120, 168}
\definecolor{model_theta_color}{RGB}{240,143,143}

\newif\ifnotes
\notestrue 

\ifdefined\counterwithin
\else
   \usepackage{chngcntr}
\fi

\newcommand{\biasColorText}[1]{{\color{bias_color}{#1}}}
\newcommand{\precisionColorText}[1]{{\color{precision_color}{#1}}}
\newcommand{\targetColorText}[1]{{\color{target_color}{#1}}}
\newcommand{\responseColorText}[1]{{\color{response_color}{#1}}}
\newcommand{\modelResponseColorText}[1]{{\color{model_responses_color}{#1}}}

\newcommand{\modelNormalColorText}[1]{{\color{model_normal_color}{#1}}}

\newcommand{\biasFormattedText}[1]{\fakemedium{\biasColorText{#1}}}
\newcommand{\precisionFormattedText}[1]{\fakemedium{\precisionColorText{#1}}}
\newcommand{\targetFormattedText}[1]{\fakemedium{\targetColorText{#1}}}
\newcommand{\responseFormattedText}[1]{\fakemedium{\responseColorText{#1}}}
\newcommand{\modelResponseFormattedText}[1]{\fakemedium{\modelResponseColorText{#1}}}

\newcommand{\modelNormalFormattedText}[1]{\fakemedium{\modelNormalColorText{#1}}}

\newtcbox{\biasBox}{nobeforeafter,colframe=bias_color!10,colback=bias_color!10,boxrule=0pt,arc=0pt, fontupper=\color{bias_color}, before upper={\rule[-2.5pt]{0pt}{10pt}},boxsep=0pt,left=1.5pt,right=1.5pt,top=0pt,bottom=0pt,tcbox raise base}

\newtcbox{\precisionBox}{nobeforeafter,colframe=precision_color!10,colback=precision_color!10,boxrule=0pt,arc=0pt, fontupper=\color{precision_color}, before upper={\rule[-2.5pt]{0pt}{10pt}},boxsep=0pt,left=1.5pt,right=1.5pt,top=0pt,bottom=0pt,tcbox raise base}

\newtcbox{\thetaBox}{nobeforeafter,colframe=model_theta_color!10,colback=model_theta_color!10,boxrule=0pt,arc=0pt, fontupper=\color{model_theta_color}, before upper={\rule[-2.5pt]{0pt}{10pt}},boxsep=0pt,left=1.5pt,right=1.5pt,top=0pt,bottom=0pt,tcbox raise base}

\newtcbox{\nuBox}{nobeforeafter,colframe=model_nu_color!10,colback=model_nu_color!10,boxrule=0pt,arc=0pt, fontupper=\color{model_nu_color}, before upper={\rule[-2.5pt]{0pt}{10pt}},boxsep=0pt,left=1.5pt,right=1.5pt,top=0pt,bottom=0pt,tcbox raise base}

\newtcbox{\normalBox}{nobeforeafter,colframe=model_normal_color!10,colback=model_normal_color!10,boxrule=0pt,arc=0pt, fontupper=\color{model_normal_color}, before upper={\rule[-2.5pt]{0pt}{10pt}},boxsep=0pt,left=1.5pt,right=1.5pt,top=0pt,bottom=0pt,tcbox raise base}

\newtcbox{\responseBox}{nobeforeafter,colframe=model_responses_color!10,colback=model_responses_color!10,boxrule=0pt,arc=0pt, fontupper=\color{model_responses_color}, before upper={\rule[-2.5pt]{0pt}{10pt}},boxsep=0pt,left=1.5pt,right=1.5pt,top=0pt,bottom=0pt,tcbox raise base}

\newcommand{\mylinesC}[2]{{\fontsize{8}{8}\color{#1}\fontfamily{qhv}\selectfont{lines #2}}\hspace{.6mm}}
\newcommand{\myColorLine}[2]{{\fontsize{8}{8}\color{#1}\fontfamily{qhv}\selectfont{line #2}}\hspace{.6mm}}
\newcommand{\mytag}[1]{{\fontsize{8}{8}\fontfamily{qhv}\selectfont\fakebook{#1}}}

\newcommand{\condition}[1]{\mbox{\selectfont\fontsize{8}{8}\fontfamily{qhv}\selectfont{#1}}\@\xspace} 
\newcommand{\variable}[1]{\textit{#1}\@\xspace}

\newcommand{\paragraphhead}[1]{\noindent{\fontfamily{qhv}{\fontsize{8}{11.5}\selectfont{\textbf{#1}}}\hspace*{5pt}\@\xspace}} 
\newcommand{\addmorevspace}{\vspace*{4pt}}
\newcommand{\addlessvspace}{\vspace*{2pt}}
\newcommand{\takeoutvspace}{\vspace*{-3pt}}

\newcommand{\pkg}[1]{{\fontsize{8.5}{11.5}\fontfamily{cmtt}\selectfont{#1}}}

\newcommand{\ie}{\mbox{i.e.,}\hspace*{-0.3mm}\@\xspace}
\newcommand{\eg}{\mbox{e.g.,}\hspace*{-0.3mm}\@\xspace}

\usepackage{pifont}

\newtcbox{\mylabel}{nobeforeafter,colframe=bgcolor,colback=bgcolor,boxrule=0pt,arc=0pt, fontupper=\color{black}, before upper={\rule[-2.5pt]{0pt}{10pt}},boxsep=0pt,left=1.5pt,right=1.5pt,top=0pt,bottom=0pt,tcbox raise base}

\newtcbox{\taglabelbg}{nobeforeafter,colframe=bgcolor,colback=bgcolor,boxrule=0pt,arc=0pt, fontupper=\color{black}, before upper={\rule[-2.5pt]{0pt}{10pt}},boxsep=0pt,left=1.5pt,right=1.5pt,top=0pt,bottom=0pt,tcbox raise base}

\makeatletter
 \def\SOUL@hlpreamble{%
 \setul{}{3ex}
 \let\SOUL@stcolor\SOUL@hlcolor
 \SOUL@stpreamble
 }
\makeatother

\usepackage{microtype}

\usepackage{xfakebold}
\newcommand{\fbseries}{\unskip\setBold\aftergroup\unsetBold\aftergroup\ignorespaces}
\makeatletter
\newcommand{\setBoldness}[1]{\def\fake@bold{#1}}
\makeatother

\newcommand{\fakebook}[1]{\mbox{#1\hspace{0.01mm}\llap{#1}}}
\newcommand{\fakemedium}[1]{\mbox{#1\hspace{0.06mm}\llap{#1}}}

\usepackage{setspace}

\usepackage[switch]{lineno} 

\usepackage{multirow,supertabular,longtable,booktabs, stackengine}

\newcounter{xReviewerTwoCounter}

\newcounter{xReviewerThreeCounter}

\newcounter{xReviewerFourCounter}

\newcounter{xReviewerFiveCounter}

\usepackage{marginnote}
\usepackage{soul}
\setul{.5ex}{1pt}
\setulcolor{lightgray}
\makeatletter
 \def\SOUL@hlpreamble{%
 \setul{}{1.9ex}
 \let\SOUL@stcolor\SOUL@hlcolor
 \SOUL@stpreamble
 }
\makeatother
\definecolor{primarycolor}{HTML}{d46373}
\definecolor{rtwocolor}{HTML}{729e9e}
\definecolor{rthreecolor}{HTML}{5b5dab}
\definecolor{rfourcolor}{HTML}{c47d41}
\definecolor{rfivecolor}{HTML}{cc6eb6}

\newtcbox{\ReviewerBox}{nobeforeafter,colframe=gray!10,colback=gray!10,boxrule=0pt,arc=0pt, fontupper=\color{gray}, before upper={\rule[-2.5pt]{0pt}{10pt}},boxsep=0pt,left=1.5pt,right=1.5pt,top=0pt,bottom=0pt,tcbox raise base}

\newtcbox{\ReviewerPrimaryBox}{nobeforeafter,colframe=primarycolor!10,colback=primarycolor!10,boxrule=0pt,arc=0pt, fontupper=\color{primarycolor}, before upper={\rule[-2.5pt]{0pt}{10pt}},boxsep=0pt,left=1.5pt,right=1.5pt,top=0pt,bottom=0pt,tcbox raise base}

\newtcbox{\ReviewerTwoBox}{nobeforeafter,colframe=rtwocolor!10,colback=rtwocolor!10,boxrule=0pt,arc=0pt, fontupper=\color{rtwocolor}, before upper={\rule[-2.5pt]{0pt}{10pt}},boxsep=0pt,left=1.5pt,right=1.5pt,top=0pt,bottom=0pt,tcbox raise base}

\newtcbox{\ReviewerThreeBox}{nobeforeafter,colframe=rthreecolor!10,colback=rthreecolor!10,boxrule=.5pt,arc=0pt, fontupper=\color{rthreecolor}, before upper={\rule[-2.5pt]{0pt}{10pt}},boxsep=0pt,left=1.5pt,right=1.5pt,top=0pt,bottom=0pt,tcbox raise base}

\newtcbox{\ReviewerFourBox}{nobeforeafter,colframe=rfourcolor!10,colback=rfourcolor!10,boxrule=.5pt,arc=0pt, fontupper=\color{rfourcolor}, before upper={\rule[-2.5pt]{0pt}{10pt}},boxsep=0pt,left=1.5pt,right=1.5pt,top=0pt,bottom=0pt,tcbox raise base}

\newtcbox{\ReviewerFiveBox}{nobeforeafter,colframe=rfivecolor!10,colback=rfivecolor!10,boxrule=.5pt,arc=0pt, fontupper=\color{rfivecolor}, before upper={\rule[-2.5pt]{0pt}{10pt}},boxsep=0pt,left=1.5pt,right=1.5pt,top=0pt,bottom=0pt,tcbox raise base}

\usepackage[bookmarks,backref=true,linkcolor=black]{hyperref} 
\hypersetup{
  pdfauthor = {},
  pdftitle = {},
  pdfsubject = {},
  pdfkeywords = {},
  colorlinks=true,
  linkcolor= black,
  citecolor= black,
  pageanchor=true,
  urlcolor = black,
  plainpages = false,
  linktocpage
}

\onlineid{1398}

\vgtccategory{Research}

\vgtcinsertpkg

\newcommand{\thetitle}{ 
Rethinking the Ranks of Visual Channels
}

\title{\thetitle}

\author{Caitlyn M. McColeman\textsuperscript{*}, Fumeng Yang\textsuperscript{*}, Timothy F. Brady, and Steven Franconeri}
\authorfooter{
\item
 Caitlyn McColeman\textsuperscript{*} is with Northwestern University. 
 E-mail: caitlyn.mccoleman@gmail.com.
 \item
 Fumeng Yang\textsuperscript{*} is with Brown University. E-mail:fy@brown.edu. 
 \newline * Both authors contributed equally to this work.
 \item
 Timothy Brady is with University of San Diego. 
 E-mail:timbrady@ucsd.edu.
 \item
 Steven Franconeri is with Northwestern University. 
 E-mail:franconeri@northwestern.edu.
}

\shortauthortitle{McColeman \MakeLowercase{\textit{et al.}}: \variable{Set Size}, \variable{Value} and \variable{Encoding}}

\abstract{%
Data can be visually represented using visual channels like position, length or luminance. An existing ranking of these visual channels is based on how accurately participants could report the ratio between two depicted values. There is an assumption that this ranking should hold for different tasks and for different numbers of marks. However, there is surprisingly little existing work that tests this assumption, especially given that visually computing ratios is relatively unimportant in real-world visualizations, compared to seeing, remembering, and comparing trends and motifs, across displays that almost universally depict more than two values.

To simulate the information extracted from a glance at a visualization, we instead asked participants to immediately reproduce a set of values from memory after they were shown the visualization. These values could be shown in a bar graph (\mbox{position (bar)}), line graph (\mbox{position (line)}), heat map (luminance), bubble chart (area), misaligned bar graph (length), or `wind map' (angle). With a Bayesian multilevel modeling approach, we show how the rank positions of visual channels shift across different numbers of marks (2, 4 or 8) and for bias, precision, and error measures. The ranking did not hold, even for reproductions of only 2 marks, and the new probabilistic ranking was highly inconsistent for reproductions of different numbers of marks. Other factors besides channel choice had an order of magnitude more influence on performance, such as the number of values in the series (\eg more marks led to larger errors), or the value of each mark (\eg small values were systematically overestimated). Every visual channel was worse for displays with 8 marks than 4, consistent with established limits on visual memory. These results point to the need for a body of empirical studies that move beyond two-value ratio judgments as a baseline for reliably ranking the quality of a visual channel, including testing new tasks (detection of trends or motifs), timescales (immediate computation, or later comparison), and the number of values (from a handful, to thousands). 

} 


\CCScatlist{ 
 \CCScat{K.number}{Category1}
}
 \teaser{
 \centering
 \includegraphics[width=.95\textwidth]{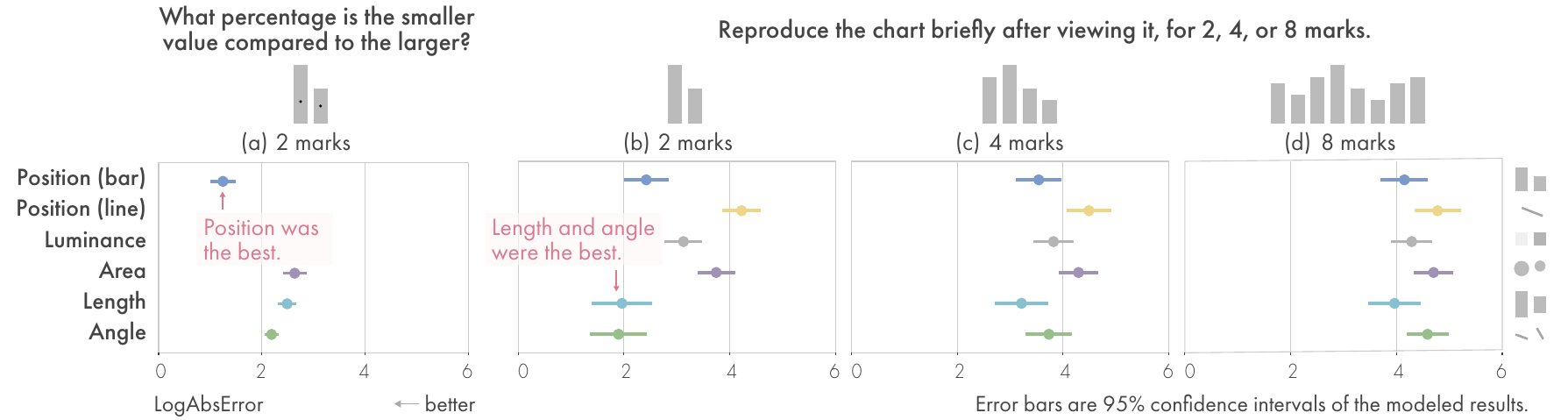}
 \vspace*{-10pt}
 \caption{
 One core guideline for data visualization design is that some visual channels offer better perceptual precision than others, drawing those precision estimates from two-value ratio judgment tasks~\cite{Cleveland1984}. 
  (a) This figure depicts typical data (from~\cite{Heer2010},  50 participants) showing these judgments are more precise for position (\eg bar graphs) than for area (\eg bubble charts).  We tested whether that ranking generalizes to the new task of reproducing 2 to 8 previously seen values, and analyzed reproduction bias, precision, and error using a Bayesian modeling approach. (b) This figure shows our modeled results (49 participants). The ranking did not hold, and other factors besides channel choice---like the number of values in the series---had an order of magnitude more influence on performance. 
 }
 \vspace*{-5pt}
 \label{fig:teaser}
}

\includecomment{comment}
\begin{document}

  
\firstsection{Introduction}
\label{sec:background}
\maketitle

Metric values can be efficiently transmitted to the human visual system across a set of channels, including position, length, or intensity \cite{bertin1983semiology} (see~\cite{munzner2014visualization} for review). When creating a visualization, designers face a choice of which channel to depict metric values, with a major constraint being a ranking of putative \textit{perceptual precision} of that channel. This ranking is organized by either expert judgment~\cite{Mackinlay1986} or operationalized by a particular task. The most referenced operationalization is the precision of making ratio judgments between two values ~\cite{Cleveland1984, Heer2010, harrison2013influencing, Talbot2014FourEO}.

\begin{figure}[b!]
\centering
\vspace{-15pt}
\includegraphics[width=\columnwidth]{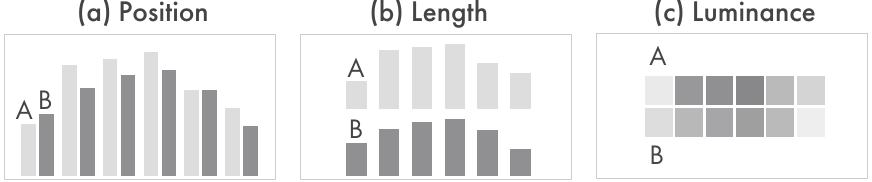}
\vspace{-20pt}
\caption{\textbf{Examples of visualization designs that use three different visual channels.} (a) This bar chart relies on the position channel for comparison, (b) this bar chart relies on the length channel for vertical  comparisons between A and B, and (c) this heatmap relies on the luminance channel. A two-value ratio judgment is precise in (a), and progressively less precise from (b) to (c).
}
\vspace{-15pt}
\label{fig:recall}
\end{figure}

For example, in Fig.~\ref{fig:recall}a, the viewer might use the position channel to estimate that the value for A is 85\% of the value for B, close to the correct answer of 80\%. The typical (log) error for this judgment is shown in Fig.~\ref{fig:teaser}a. It is typically the lowest error of any channel. Making the same judgment in Fig.~\ref{fig:recall}b between A and B (now separated vertically) is a bit tougher, as reflected by the larger error value for ratio judgments of length in Fig.~\ref{fig:teaser}a. 
Finally, Fig.~\ref{fig:recall}c shows the same data plotted as luminances. While we do not know of empirical measurements of ratio judgment error for this channel, expert judgment \cite{Cleveland1984} (as well as ours) suggests that the error would be quite high~\cite{Mackinlay1986}.

\subsection{Beyond a ranking based on two-value ratio precision}
\label{sec:beyond-twovalues}

The channel ranking derived from error measurements of two-value ratio judgments likely deserves its role as a key factor that determines the choice of a channel for depicting metric values. But there is an implicit assumption that it should extrapolate broadly across the types of lower-level visual tasks that viewers execute in real-world visualizations and visual analytics. This is a bold assumption, because visualizations require that we extract, remember, and compare sets of statistics, trends and motifs, across visualizations that almost universally depict more than two values,  leading to increasing unease about the dominance of this method of ranking channels \cite{bertini2020shouldn}. A taxonomy of such operations presents ten low-level perceptual tasks used in analyzing a set of datasets~\cite{Amar2005}. 
Interestingly, computing ratios does not appear as a task. `Retrieving a value’ is present, and it is plausible that this task is the foundation of a two-value ratio judgment for charts\cite{Cleveland1984}. 
A more recent survey includes `computing derived value` but also reveals concerns about task-dependent effectiveness~\cite{quadri2021vistasks}. 
In Fig.~\ref{fig:recall}a, if the viewer knew that the maximum value of the $y$-axis were 10, computing a ratio between any bar and that number would allow the viewer to extract the value of a single bar from an unlabeled (or sparsely-labeled) axis. 

In Fig.~\ref{fig:recall}, for each of the three visualizations, where are the three highest, or lowest, values for A or B? Where are the largest (or average) differences for each value pairing across the series? There are dozens of such critical comparisons that all involve more than two points (see~\cite{gleicher2011visual, bertini2020shouldn, nothelfer2019measures} for review), and there is insufficient empirical work that evaluates whether the ranking of channels extracted from two-value ratio tasks also applies to them (see Sec.~\ref{sec:related-work}).

\subsection{The present study: reproduction as a proxy for various comparison tasks}
\label{sec:present-study}

How might one compare performance for each channel across such a long list of potential comparison tasks? 
We start with the assumption that many of these comparison tasks require that one set of values be held in visual memory, and that memory is compared to a subsequently perceived set. For example, in any panel of Fig.~\ref{fig:recall}, computing a two-value ratio might not feel like it requires a heavy memory component. But comparing the global shape of series A versus series B feels far more capacity-limited \cite{trick1997clusters} and memory-intensive~\cite{franconeri2013nature}. Indeed, evidence from the visual memory and attention literatures suggest that for such more complex comparisons, one must first inspect A, hold 
the set in memory, and then compare that memory to set B~\cite{yu2019gestalt, yu2019similarity, huang2007boolean}. At the very least, comparisons that are not `within the eyespan' \cite{tufte1990envisioning}, requiring an eye movement or turn of a page, certainly require, and will be limited by, visual memory. 

Visual memory is highly capacity-limited \cite{schurgin2020psychophysical}. As we attempt to remember more information, precision plummets, bias quickly increases, and storage capacity hits ceiling limits (see Sec.~\ref{sec:visual-memory}). Therefore, we would expect the number of data values involved in comparison tasks to predict whether the viewer is successful. Because memory serves as a critical gateway for performance in comparison tasks, the present study measures how a viewer’s memory precision, bias, and overall error is affected by the channel used to encode a dataset, and how those measures are affected by the number of data values that the viewer is asked to process and remember.

The present study measures memory using a reproduction task, under the assumption that this measurement will generalize to a variety of comparison tasks. 
If we had instead used a more specific comparison task, which would we pick? Comparisons of data distributions? A search for the longest set of relatively low values? Ask for the differences in the global shape across the two series? If so, what type of difference, and how would it be reported? And how different should the two data series be, and in what ways? The present reproduction task allows a first look at how channel and number of marks affects reproduction performance, without the need to consider these more specific operationalizations of the various types of visual comparisons. We hope that after this initial exploration, the field can begin to ask more targeted empirical questions for particular comparison tasks.  

We asked participants to immediately reproduce a set of values seen moments earlier across six channels and three numbers of marks \{2, 4, 8\}. 
Our results from a Bayesian multilevel model show that the previous ranking~\cite{Cleveland1986} does not hold, even for reproducing only 2 marks. The new probabilistic ranking also varies with the number of marks. Other factors besides channel choice have an order of magnitude more influence on performance, such as the number of marks in the series, or the value of each mark. Across every visual channel, performance drops precipitously when more than just a few marks have to be stored,
consistent with the known limits on visual memory.

\subsection{Contributions}
\label{sec:contributions}

This work challenges the assumption that the ranking derived from the precision of judging a ratio between two visual marks will extrapolate to new tasks, especially those that involve more than two marks. 

\vspace{2pt}

\noindent Our primary contributions are as follows.
	\begin{itemize}[leftmargin =10pt, topsep = 2pt, itemsep = -1pt]
	
    \item \textbf{Experimental study results} on the effects of six typical encoding channels, and the number of marks \{2, 4, 8\}, on a task of reproducing a set of visualized data, leading to a reassessment of the value of rankings based on two-value ratio tasks.  
   
	\item  \textbf{A contextual, probabilistic ranking} of the six visual channels on three statistical measures: bias, precision, and error. 
	
	\item \textbf{A publicly-accessible dataset} of 28,602 responses measuring that reproduction performance, as well as a Bayesian multilevel  model to describe the dataset. 
	   The dataset, analysis script, and model files are available at {\bf{\pkg{\selectfont\fontsize{7}{8}\href{https://doi.org/10.17605/OSF.IO/3E2QT}{https://doi.org/10.17605/OSF.IO/3E2QT}}}}.
	
	\end{itemize}

\section{Related work}
\label{sec:related-work}
Here we surveyed work in visual perception, information visualization, and visual working memory to gather considerations for factors that may impact visual reproduction performance.

\subsection{Context and bias effects on visual judgements}

While Cleveland and McGill \cite{Cleveland1986} tested the precision of ratio judgements with only two relevant values for the judgment itself, they also showed decreased precision for displays where those values were crowded by adding other values in the display \cite{Cleveland1984, Heer2010}.
More recent work~\cite{zhao2019neighborhood,Talbot2014FourEO} identified similar impairments. 
In other reproduction tasks, like the one used in the present study, surrounding values in a display created memory biases, such that recollections of a single relevant value were repulsed from the 0, .5, and 1.0 proportion of a second larger reference bar \cite{mccoleman:2020:Bias}. Memory bias has been shown even for values presented alone, such that tall bars with a high height:width ratio were underestimated, and wide bars with a low height:width ratio were overestimated~\cite{CCC:2020:Bars}. 


\subsection{Evaluations beyond two-value ratio precision}
After one study showed that correlation judgments follow a systematically measurable profile of perceptual precision for scatterplots using the position channel \cite{rensink2010perception}, a later study ranked the relative precision of correlation judgments across other visualizations, finding that position-based scatterplots offered the highest precision, but position-based line charts  offered the lowest precision~\cite{harrison2014ranking}. Angle, a channel with low precision on a two-value ratio task, showed the second-highest precision~\cite{harrison2014ranking}. Though in this case, the correlation judgment may not have been perceptually extracted by angle \textit{per se}, but emergent shapes created by the angles for high negative correlations. 

\begin{figure*}[t!]
 \centering
  \vspace*{-2pt}
 \includegraphics[width=\textwidth]{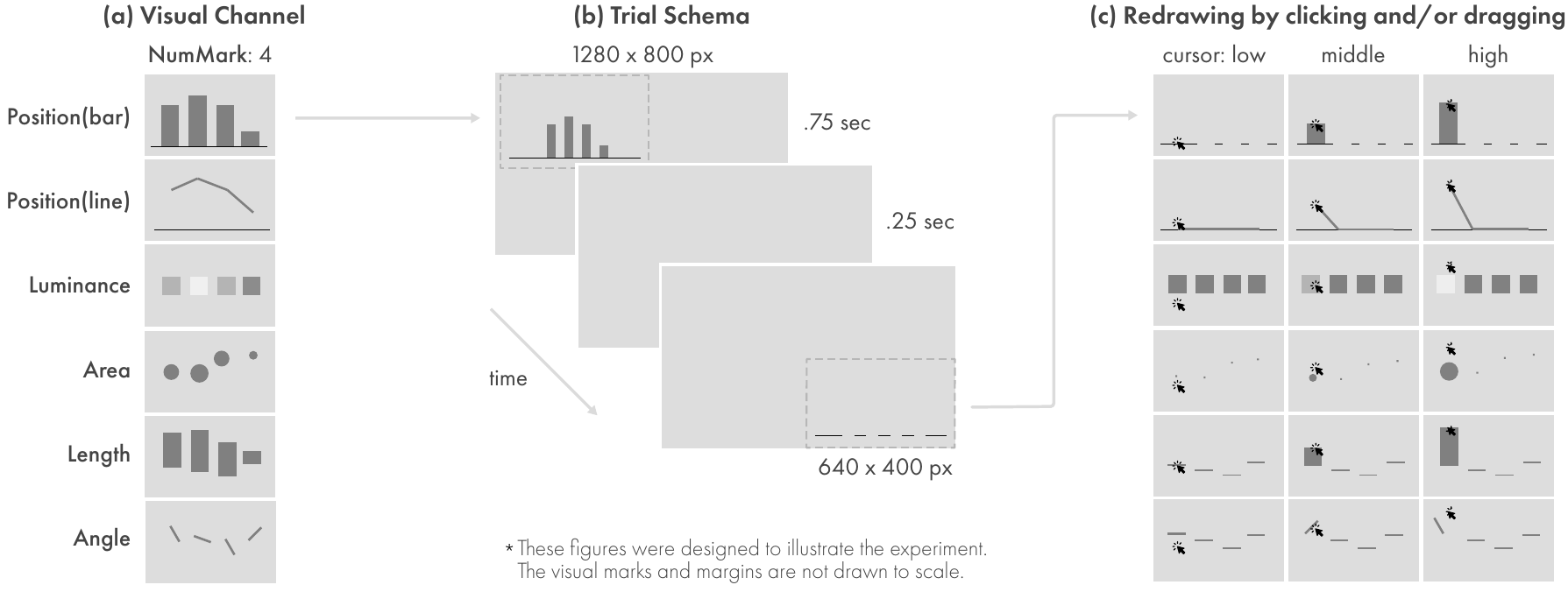}
 \vspace*{-15pt}
 \caption{\textbf{Visual channels and the reproduction task}.
 (a) Examples of visual channels for  4  visual marks. 
 (b) Each trial followed a ``show-remove-reproduce'' procedure to indicate their responses. 
 (c) In reproduction, participants clicked on the screen or dragged the mouse to redraw the previously-seen marks. In all conditions, the visual channel changed as a linear function of the vertical position of the mouse cursor, such that even \condition{angle} and \condition{area} were changed by dragging the mouse up-and-down. For \condition{area} channel, participants adjusted based on area not radius.
More details are available in Appendix~A.
}
 \vspace*{-9pt}
\label{fig:channelRedrawing}
\end{figure*}

With judgments of aggregate properties of mean, average, or spread, the typical ranking can reverse, such that typically low-ranked values like luminance (in this case, a ramp combining luminance and color saturation) can actually lead to the best performance in those tasks \cite{albers2014task} (see \cite{szafir2016four} for review). Judgments of minimum, maximum, or range were still best for visualizations that used position channels. Another study asked participants to complete four tasks---read value, compare values, find maximum, and compare averages---across visualizations that relied on position, size, or color (similar to the luminance ramp used here). They found similar results, where extracting one value, or comparing two single values, was fast and accurate for position, but for aggregate properties like comparing averages, the color condition showed equal performance \cite{kim2018assessing}.
Another study, similar in spirit, tested the speed, accuracy, and preference for ten data visualization tasks across scatterplots, bar charts, pie charts, and line charts \cite{saket2018task}. They found, for example, advantages for bar charts in finding value clusters, or that scatterplots show advantages for anomaly detection, but not for cluster detection. 

Others evaluated the visual channels for comparison (measured by staircasing threshold differences that could be detected in a limited time \cite{yuan2019perceptual}) across two tasks, finding the maximum difference among two paired values in a display similar to the left bar chart in Fig.~\ref{fig:recall}, or the stronger correlation between two such pairings of values. The study included bar, line, and donut charts, was focused on comparing value arrangements within each chart type (\eg juxtaposed vs. interleaved values). Those charts---and their underlying channels---could in theory be compared in their effectiveness for supporting those comparison tasks, but differences in the methods between chart types make that comparison difficult~\cite{ondov2018face}.

Similar to the cited studies, the present study relies on a single task, but we regard reproduction as a starting point for more generalizable results, compared to two-value ratio precision or a single visual comparison task.

\subsection{Visual memory}
\label{sec:visual-memory}


 Working memory is the ability to hold information actively in mind, and to manipulate that information to perform a wide variety of cognitive tasks~\cite{baddeley2001concept}. For visual memory in particular, when asked to remember visual information across eye movements (\eg for comparisons) or across interruptions~\cite{hollingworth2008understanding}, studies typically claim a capacity limit of only `3-4 items’ (\eg~\cite{cowan2001magical}). 
Even for fewer than 3-4 items, when participants recall the sizes, colors, or angles, of previously seen objects, they are notably less accurate in recalling 2 items than 1 item (\eg~\cite{zhang2008discrete, bays2009}). 

Remembering more complex conjunctions of visual channels (\eg both the color and orientation of a mark) is extremely difficult when more than 1-2 objects must be remembered~\cite{oberauer2013visual,hardman2015remembering}. 
The performance cost of increasing memory load from just 1 item held in mind at once to 2 items is larger than the cost of increasing the load from 4 to 8 items (\eg~\cite{schurgin2020psychophysical}). 
Thus, the profile of memory performance for tasks that involve only 1 or 2 items at a time may not predict the profile for more complex visual displays~\cite{brady2015no}. 
There are also strong contextual dependency effects where values are stored in compressed ways, as relative to other values ~\cite{brady2015contextual}. 
In a visualization, increasing the number of memorized values will lead to performance changes that are hard to predict. Since nearly all data visualizations include more than 1 or 2 marks, it is critical to study these cases directly rather than assume the lessons drawn from studies of 1 or 2 marks will generalize to these larger value sets. 

In the present study, participants were asked to reproduce data displays that fall \textit{within} (2 marks), \textit{at} (4 marks), or \textit{beyond} working memory capacity (8 marks) to gather data from qualitatively different memory loads. 
Participants in this task rely on reproduction of values, as opposed to semantic recall of the main message of a visualization~\cite{kong2019trust} or whether they have encountered an entire image before \cite{borkin2015beyond}.   
This task is an analogy to typical visual working memory tasks, acting as a proxy for how one retains values of marks across eye movements and delays (as when reading the text associated with the visualization). 

\section{Methods}
\label{sec:methods}

This section presents and justifies our design decisions, along with the description of the stimuli generation process, the experimental design and procedure, and the data collected.

\subsection{Visual channels}

%
As introduced above, we chose six visual channels (denoted by \variable{VisualChannel}) to cover a wide range of the original ranks by Cleveland and McGill~\cite{Cleveland1986}:
\condition{\mbox{position (bar)}} (bar chart), \condition{\mbox{position (line)}} (line chart),  \condition{luminance} (heatmap), \condition{area} (bubble chart), \condition{length} (misaligned bar chart), and \condition{angle} (wind map).
We show an example of each of the six visual channels in Fig.~\ref{fig:channelRedrawing}a.

\begin{figure}[b!]
 \centering
 \vspace{-12pt}
 \includegraphics[width=\columnwidth]{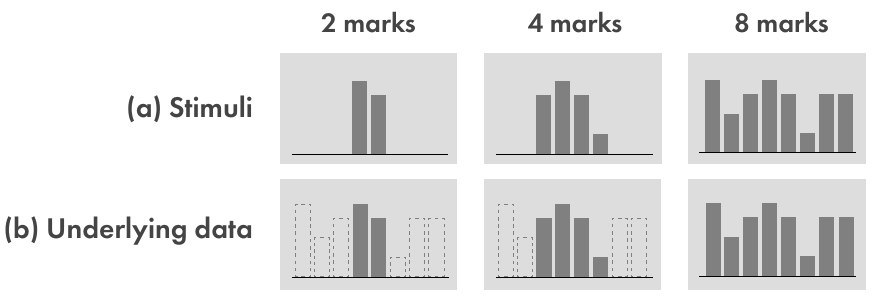}
 \vspace{-18pt}
 \caption{\textbf{Different numbers of visual marks}. 
 We used the same pre-generated datasets across different \variable{NumMark} and \variable{VisualChannel} and  removed the side values when showing  2 or 4 marks.}
 \label{fig:setsizeDiff}
\end{figure}

\subsection{The number of marks}
We tested three different numbers of marks (denoted by \variable{NumMark}): 2, 4, and 8 (Fig.~\ref{fig:setsizeDiff}a). 
The 2-mark condition requires that the viewer extract the value of two data visualization marks, replicating the earlier studies based on  
two-value ratio judgments (\eg~\cite{Cleveland1984, Heer2010, harrison2013influencing, Talbot2014FourEO}). The difference in our task is the nature of that extraction, in that participants must redraw it rather than reporting a ratio value. The 4-mark condition aligns with the boundary of typical working memory capacity, and the 8-mark condition exceeds even the most optimistic estimates for human visual working memory.
These three conditions have categorically different loads for working memory, allowing us to infer how the working memory limits affect reproduction. 

\subsection{Experimental design}
We split the visual channels into two experiments based on whether the channel uses a common baseline. 
This split was decided to align participants' mental models and to keep the experiment duration approximately 30 minutes to avoid severe fatigue effects. 
The first experiment tested \condition{\mbox{position (bar)}}, \condition{\mbox{position (line)}}, and \condition{luminance}. 
The second experiment tested \condition{length}, \condition{area}, and \condition{angle}. 
Each experiment tested all three numbers of marks \{2, 4, 8\}.
Each participant did the task with 3 visual channels and all 3 numbers of marks, but with different channels for different experiments.

Each pair of \variable{VisualChannel} $\times$ \variable{NumMark} was a block with a series of trials. 
The first experiment used 13 trials per block.
The second experiment used 15 trials per block; this is because, in the pilot study, we found that the second experiment was more difficult: the mapping between the vertical mouse click and the visual change was challenging, and responses were noisier.
Thus, we included the additional two trials to offset this additional noise.
Within each of the two experiments, the order of visual channels was counterbalanced.


\subsection{Generating stimuli}
All the values were in the numeric range of [0.01, 1.0] and encoded to the visual channels as follows (see Appendix A for more details).
The dimensions of marks were decided to maximize the varying range but to avoid overlapping.
The background was set to \pkg{rgb(.75,.75,.75)} (light grey) to control visual contrast effects.
As a result, \condition{\mbox{position (bar)}} has the height of each bar ranging from 3.9 pixels to 390 pixels.
\condition{\mbox{Position (line)}} has the height of each line end ranging from 3.9 pixels to 390 pixels.
\condition{Luminance} has the color of each square ranging from \pkg{rgb(.5,.5,.5)} (grey) to \pkg{rgb(1.0,1.0,1.0)}(white) such that its middle point was the same as the background color.
\condition{Area} has the area of each circle ranging from $\pi(5 + 1.19\ \textrm{pixels})^2$ to $\pi(5 + 37.5\ \textrm{pixels})^2$; the 5 pixels offset was to ensure that all the circles were visible all the time.
\condition{Length} has the height of each bar ranging from 3.75 pixels to 375 pixels.
Lastly, \condition{angle} has each segment rotated counter-clockwise in the range of 1.8$\degree$ to 180$\degree$.
For \condition{area}, \condition{length}, and \condition{angle}, the vertical position of the marks were randomly generated in the range of the $y$-axis, spanning .0 of its height (\ie the bottom of the axis range) to .9 of its height.
 

All datasets were pre-generated, and the same datasets were repeated within the same experiment for different \variable{VisualChannel} $\times$ \variable{NumMark} blocks. 
Each dataset consisted of 8 numeric values, and each value was randomly and uniformly  sampled from the standardized values of \{0.1, 0.2, 0.3, 0.4, 0.5, 0.6, 0.7, 0.8, 0.9,  1.0\}. 
When the \variable{NumMark} was 8, participants saw all 8 values in a trial. When the number of marks was 2 (or 4), participants saw the middle two (or four) values; the remaining values were not displayed (Fig.~\ref{fig:setsizeDiff}b).
Each participant viewed different datasets within a \variable{VisualChannel} $\times$ \variable{NumMark} block, and repeated the same datasets across different blocks.
The order of the datasets and the values within a dataset were otherwise randomized.

\addlessvspace

\subsection{Procedure}
The experimenter first collected informed consent from the participants and then shared an instruction presentation displaying the format, structure and response modality for all trial conditions. 
The experimenter was present for training and answered clarifying questions the participant had about how to make their response.

\addlessvspace



\paragraphhead{Trial}
As discussed above, in each trial, participants performed a reproduction task.
They first saw the stimuli visualization  for .75 seconds. 
The stimuli were then replaced with a blank screen for .25 seconds.
Immediately after this, participants were asked to reproduce each visual element (\eg a bar) as they clicked and/or dragged the mouse to change the pre-marked visual elements on the screen (Fig.~\ref{fig:channelRedrawing}c).
The stimuli visualization was randomly placed in one of the four quadrants (Fig.~\ref{fig:channelRedrawing}b) and redrawn in the diagonal quadrant.
For example, if participants saw the stimuli in the upper left, they redraw the stimuli in the bottom right. 

The short duration exposure, along with unlabeled axes, prevent participants from recoding stimuli into other forms~\cite{mccarthy1990shortterm} and suppress top-down effects like prior knowledge. The duration is adequate for testing visual working memory~\cite{mccarthy1990shortterm} and provides ample  time  for  the  vision system to encode information (\eg comparing correlation in scatterplots~\cite{rensink2014prospects}, estimating two-value ratio in bar charts~\cite{mccoleman:2020:Bias}, etc.). 
The inclusion of a blank screen as a mask and a different redrawing location together eliminated visual aftereffects.




\addlessvspace

\paragraphhead{Participants}
Thirty and twenty-nine participants were recruited for the two experiments, respectively. 
They were undergraduate students from the same institution, enrolled in introductory psychology classes, for which they earned partial credit in exchange for their time. 
Participants were between 18 and 23 years old ($\mu$ $=$ 19.02 years, $\sigma$ $=$ 0.96; 22 female, 34 male, 3 unspecified), all with normal, or corrected-to-normal vision.
The same author and experimenter proctored all the experiment sessions and finished them before the COVID-19 pandemic.

\addlessvspace

\paragraphhead{Apparatus}
The experimental system was implemented using Psychophysics Toolbox~\cite{brainard1997,kleiner2007} and  MATLAB 2018a, running on a Mac Mini (OS 10.10.5). Stimuli were displayed on a 23" monitor with a resolution set to 1280 $\times$ 800 pixels and a 60 Hz refresh rate. 
Participants were sat approximately 18.5" from the display. 

\subsection{Response data}
All the raw data from all the participants were considered for analysis with two exceptions.
First, 3 and 7 participants from the two experiments, respectively, contributed to the pilot study or were unable to finish the experiment; they were excluded for the purpose of balancing learning and fatigue effects. 
Second, in the \condition{angle} condition, when showing a maximum value 1.0 (180\degree) as the reference, 45.79\% of the responses were the same default value of 0.001 ($\sim$0\degree), resulting in a very large error (100\% error).
Because both 0\degree and 180\degree were  a flat segment (see Fig.~\ref{fig:channelRedrawing}), we think, if not all, the majority of the participants misinterpreted 180\degree~as 0\degree.
To ensure the comparability of our results, we transferred the reference value (1.0) to 0.0 (180\degree~to 0\degree) for  \condition{angle}.

We recorded the reproduced value of each mark, the order of visual marks,  the reference values shown on the screen,  the reaction time, \variable{VisualChannel}, \variable{NumMark}, and the trial index. 
We collected 6,129 trials
 $=$ 3 \variable{VisualChannels} $\times$ 3 \variable{NumMarks} $\times$ (13 trials $\times$ 27 participants $+$ 15 trials $\times$ 22 participants). 
Together we analyzed 28,602 responses $=$ 3  \variable{VisualChannels} $\times$ (2 $+$ 4 $+$ 8) marks $\times$  (13 trials $\times$ 27 participants $+$ 15 trials $\times$ 22 participants).


\section{Analyses}
\label{sec:analyses}

To analyze the response data, we first decided the measures to quantify the effects, followed by a description of the modeling approach and the model to support the inference.

\subsection{Measures}
\label{sec:measures}

\begin{figure}[h!]
 \centering
 \includegraphics[width=\columnwidth]{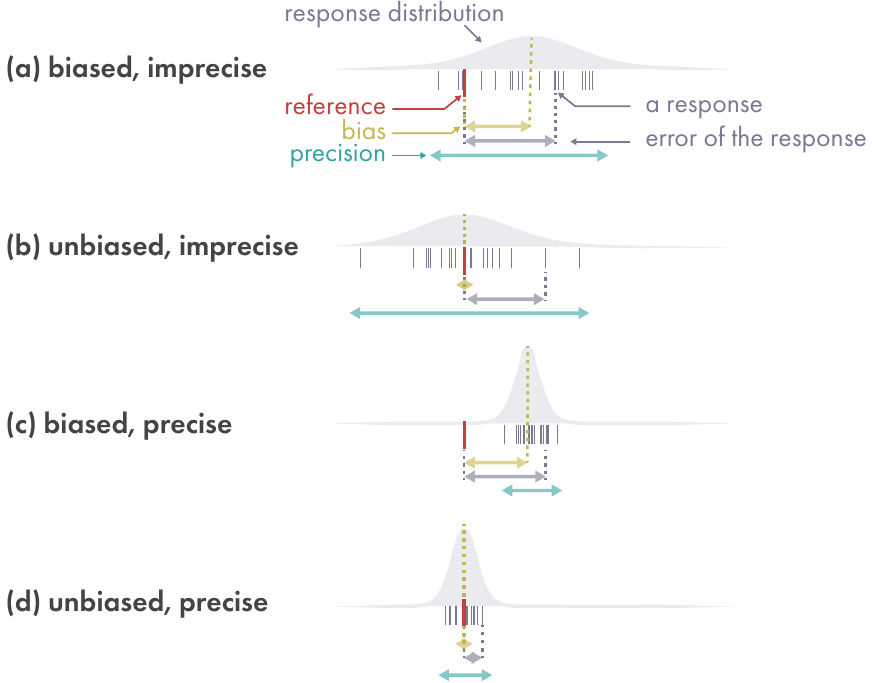}
 \vspace*{-15pt}
\caption{\textbf{\biasColorText{Bias}, \precisionColorText{precision}, and  \responseColorText{error}}.
\biasFormattedText{Bias} and \precisionFormattedText{precision} describe the average properties of a set of responses, while \responseFormattedText{error} is a measure for a single response. 
In this work, \responseFormattedText{error} is defined as the deviation from the reference,  mean of errors is defined as \biasFormattedText{bias}, and standard deviation of errors is defined as  \precisionFormattedText{precision}.}
 \vspace*{-10pt}
\label{fig:measures}
\end{figure}

We follow the literature on visual memory and used three statistical measures to compare participants' responses: \biasFormattedText{bias}, \precisionFormattedText{precision}, and individual response level \responseFormattedText{error}~\cite{brady2015contextual} (see Fig.~\ref{fig:measures}).

Among these, \biasFormattedText{bias} is how the mean of the responses deviates from the actual value presented as \targetFormattedText{the reference}. 
Think of bias as systematic error or the tendency to make mistakes in a certain direction, such as exhibiting a bias to overestimate wide bars~\cite{CCC:2020:Bars}. 
\precisionFormattedText{Precision} is the consistency of participants' responses; they may consistently report the same value, regardless of \targetFormattedText{the reference} value.
\biasFormattedText{Bias} and \precisionFormattedText{precision} are different facets for the same set of responses. 
Participants could be precise but consistently underestimate (or overestimate) the value~\cite{mccoleman:2020:Bias,Cleveland1984}. 
They could be imprecise but generally right on average.
Alternatively, \responseFormattedText{error} measures how each response deviates from the reference value.
These three measures are different facets for the same distribution of the responses, capturing variations in visual error and reproduction performance through different lenses.

Here we used a \pkg{student\textunderscore{}t(\biasFormattedText{$\mu$},\precisionFormattedText{$\sigma$},$\nu$)} distribution for a more robust understanding. 
 \biasFormattedText{Bias},  the mean of responses, is described by the location parameter \biasFormattedText{$\mu$}; and \precisionFormattedText{precision},  the consistency of responses,  is described by the dispersion parameter~\precisionFormattedText{$\sigma$}\footnote{
Strictly, the $\sigma$ parameter (standard deviation) describes \textit{imprecision}.
}. 
The \responseFormattedText{errors} of individual responses combine both \biasFormattedText{bias ($\mu$)} and  \precisionFormattedText{precision ($\sigma$)} of the responses into one measure.
If we fit the distribution with the response data collected, then knowing \biasFormattedText{$\mu$} and \precisionFormattedText{$\sigma$}, we are able to draw samples from the distribution and calculate  \responseFormattedText{error} of each draw. 

It is important to note that \biasFormattedText{bias} and \precisionFormattedText{precision} describe the average properties of a set of responses (\eg responses from one or more experimental conditions, one or more participants). 
However, \responseFormattedText{error} is a measure for a single response, combining variance from \biasFormattedText{bias} and \precisionFormattedText{precision}; hence it is  with more uncertainty than \biasFormattedText{bias} and \precisionFormattedText{precision}.

Because each of the three measures is associated with a reference, in the remainder of this paper, we subtract the reference value from each response and transform all the raw responses to \responseFormattedText{errors} (\ie relative responses $=$ raw responses $-$ reference values).

\vspace{5pt}
\subsection{Bayesian multilevel (hierarchical) modeling}
\label{sec:bayesian}

We adopted a Bayesian modeling approach to estimate the error distribution. 
The mean and standard deviation parameters of this distribution, as described above, are considered bias and precision of the responses.

We followed a process of model expansion with \textit{regularization}~\cite{mcelreath2016statistical,pu2018garden}.
It allowed us to understand how each predictor affects the model, to capture more variance in the data while reducing overfitting, and to explore the effects of secondary variables. 
We started with a \textit{minimal model}, which contained only experimental variables, and a list of potential predictors, ordered by their importance in our subjective beliefs.
We then progressively added the predictors and evaluated each intermediate model by inspecting their posterior predictions and posterior distributions of the coefficients.
We compared each intermediate model to the last model using WAIC (widely applicable information criterion) and LOO (Leave-One-Out Cross-Validation) for out-of-sample prediction accuracy, and examined their Akaike weights (the probabilities of the differences in these predictions)~\cite{mcelreath2016statistical,lambert2018student}.
We also started with weakly informative priors and gradually regularized the priors as the model expanded~\cite{mcelreath2016statistical}.
We chose the final model which was the best at addressing our research questions, describing the current data, and predicting future observations.

We implemented the modeling processes using \pkg{R} packages  
\pkg{brms}~\cite{burkner2017brms}, \pkg{CmdStanR}~\cite{cmdrstan}, \pkg{bayesplot}~\cite{bayesplot2019,bayesplot2021}, \pkg{ggdist}~\cite{kay2020ggdist}, and \pkg{tidybayes}~\cite{kay2020tidybayes}. We provide the analysis script and the resulting model files as supplementary materials (\href{https://osf.io/b26yq/}{the \pkg{analysis.Rmd|html} and \pkg{*.rds} files}).

\newpage
\subsection{Model specification}
\label{sec:model}

\paragraphhead{Formula}
Using a syntax similar to \pkg{brms}'s~\cite{burkner2017brms} extended Wilkinson-Rogers-Pinheiro-Bates notation~\cite{pinheiro2017package,wilkinson1973symbolic}, our final model is  

\begin{figure}[H]
 \centering
 \vspace{-11pt}
 \includegraphics[width=\columnwidth]{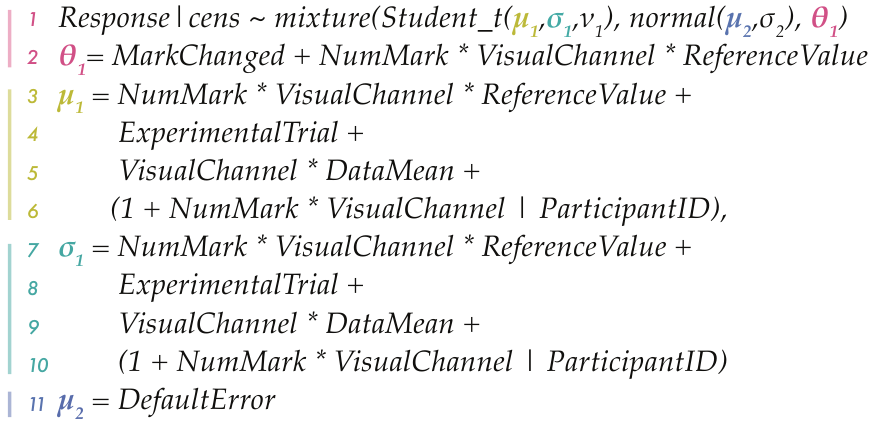}
 \vspace{-14pt}
\end{figure}
\begin{minipage}[t]{10pt}
\begin{figure}[H]
\centering
\vspace{-13pt}
\hspace*{-20.5	pt}\includegraphics[height=525pt]{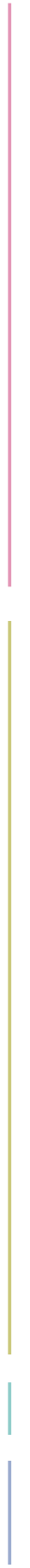}
\end{figure}
\end{minipage} 
\edef\myindent{\the\parindent}
\hspace*{-23.5pt}\begin{minipage}[t]{255pt}\setlength{\parindent}{\myindent}
\vspace{-14pt}
\hspace*{-7.5pt}\paragraphhead{Explanation}
\begin{itemize}[leftmargin=10pt, itemsep =0pt, topsep = -2pt,label={}]
\item \myColorLine{model_responses_color}{1} 
We treat all the responses  as arising from a mixture of two distributions: a \pkg{student\textunderscore{}t} distribution for all the genuine reproduction responses, and a \pkg{normal} distribution for those made without an intention to reproduce a value, termed the `default' distribution. 
This is because sometimes participants did not move the mouse to make a response, resulting in a cluster of likely irrelevant responses at a small (known) value.  
The mixture model separates these two sorts of responses; a mixture model like this is ubiquitous in the visual memory literature~\cite{zhang2008discrete,brady2015contextual} for modeling responses.

\takeoutvspace

In the model,  the mixture parameter \modelResponseFormattedText{$\theta_1$}, 
the mean (\biasFormattedText{$\mu_1$; bias}), and standard deviation (\precisionFormattedText{$\sigma_1$; precision}) of the \pkg{student\_t} distribution vary with the experimental variables.
The mean (\modelNormalFormattedText{$\mu_2$}) of the \pkg{normal} distribution  captures the default responses (see \myColorLine{model_normal_color}{11}below).
We assumed that the $\nu_1$ parameter of the \pkg{student\_t} distribution and the standard deviation ($\sigma_2$) of the \pkg{normal} distribution do not vary.  
We also left censored the responses to reduce the impact of erroneous responses.

\takeoutvspace

\myColorLine{model_responses_color}{2}
This line describes the probability of a response coming from the genuine reproduction (cf. default) distribution. 
This probability could be affected by if the mark was changed (1 or 0), the number of marks, the visual channel used, and the reference value.

\item The mean (\biasFormattedText{$\mu_1$; bias}) of the reproduction distribution is a joint function of a set of linear predictors with varying intercepts and slopes:

\takeoutvspace

\myColorLine{bias_color}{3} The experimental variables \variable{NumMark} and \variable{VisualChannel} are of the most importance.
\variable{ReferenceValue} acknowledges that perceptual
errors are likely to be affected by the magnitude of stimuli (\eg Weber-Fechner's~\cite{hecht1924visual,gescheider2013psychophysics}, Stevens's power~\cite{stevens1957psychophysical}, and Guilford's laws~\cite{guilford1932generalized}) without making a strong assumption about this relationship is the same for different numbers of marks and visual channels; this aligns with the observations that Weber's law appears not to hold for extreme values~\cite{gaydos1958sensitivity} nor perception of area and angle~\cite{stevens1957psychophysical} (see Appendix~B for more discussion).
The interaction between these variables further generalize this relationship.

\takeoutvspace

\myColorLine{bias_color}{4}  \variable{ExperimentalTrial} captures learning and fatigue effects over the course of the experiment such that we can later divest these effects by conditioning on the median trial.

\takeoutvspace

\myColorLine{bias_color}{5} \variable{DataMean} is the average of the shown data in a trial. 
It approximates the context of a response.
If the reference value is small but the data mean is large, it may indicate that this response was made in the presence of other large values, and \textit{vice versa}.
The interaction with  \variable{VisualChannel} is motivated by the speculation that participants may use perceptual proxies for mean~\cite{jardine2019perceptual,ondovb2020revealing}, and the proxies may be different for different visual channels~\cite{harrison2014ranking}.

\takeoutvspace

\myColorLine{bias_color}{6} The group-level effects (``random intercepts and slopes'') capture the correlation within a participant and also allow each participant to vary for different experiments and experimental conditions.

\item  \mylinesC{precision_color}{7-10}  The same predictors were used for \biasFormattedText{bias ($\mu_1$)} and \precisionFormattedText{precision ($\sigma_1$)} to ensure compatibility.

\item \myColorLine{model_normal_color}{11}   The responses from the default distribution, when participants may not be trying to reproduce the value, are always near a small, known value (denoted by \variable{DefaultError}), specified via the informative priors for the mean (\modelNormalFormattedText{$\mu_2$} and standard deviation  ($\sigma_2$) .
\end{itemize}
\end{minipage}

\begin{figure*}[!t]
\centering
\vspace*{-10pt}
\includegraphics[width=\textwidth]{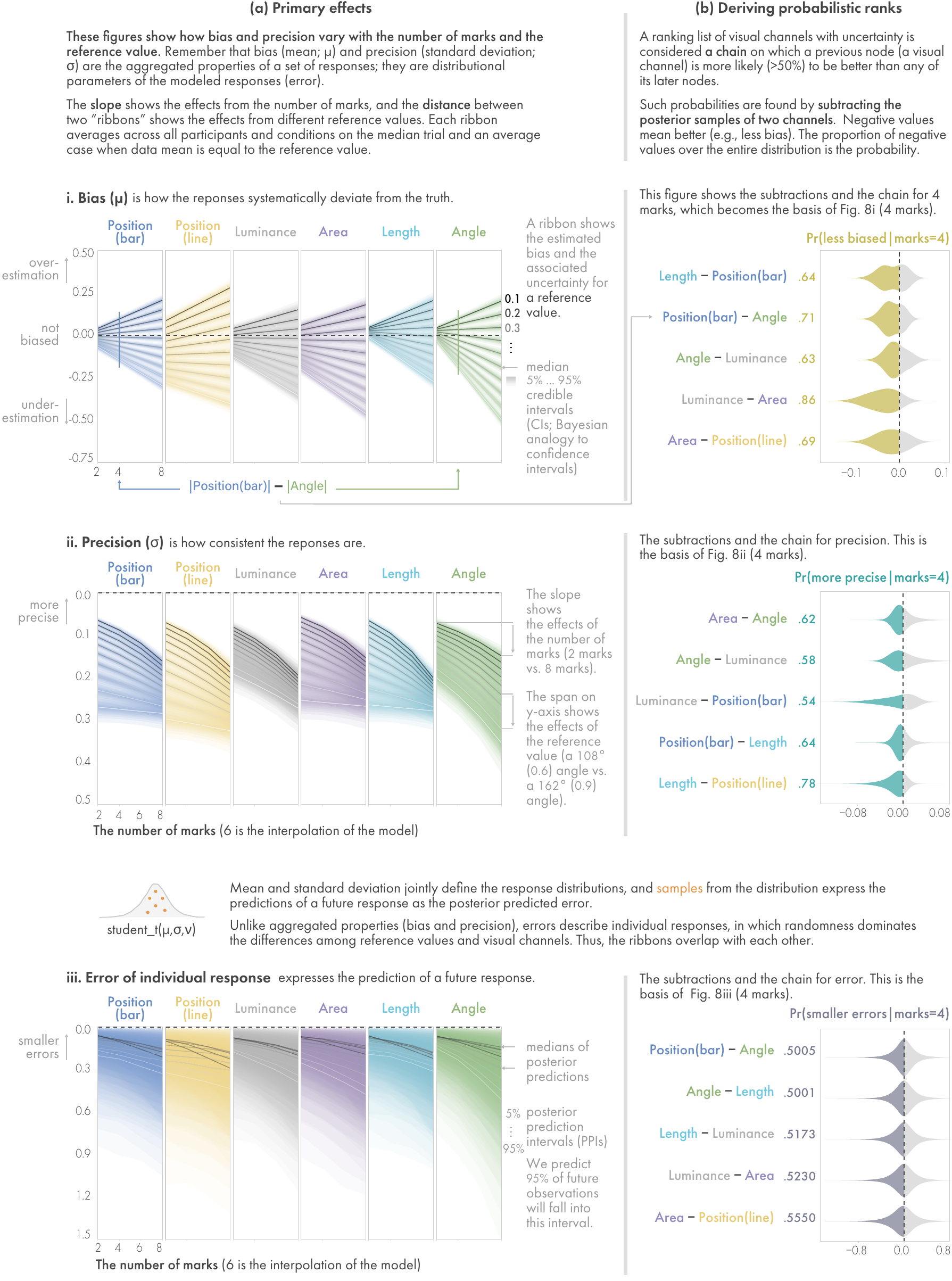}
\vspace*{-15pt}
\caption{(a) \textbf{The primary effects} modeled from the experimental observations; and (b) how we compare two visual channels, calculate the probabilities of being better, and finally \textbf{derive the probabilistic ranks.}
} 
\label{fig:primary-effects}
\vspace*{-20pt}
\end{figure*}
\section{Results}
\label{sec:results}
To understand the differences in visual channels for the reproduction task, we report various effects on each of the precision, bias, and error measures.
We then derive ranks for the visual channels.

We base our inference on the first distribution of the mixture model and the posterior distributions (marginal, conditional, and predictive distributions). 
Marginal posterior distributions summarize all the known information for one parameter; 
conditional posterior distributions tell us the expected value of one parameter in a specific situation; 
and posterior predictive distributions provide unobserved data conditioning on the observed data and the fitted model.

\begin{figure}[!ht]
\centering
\includegraphics[width=\columnwidth]{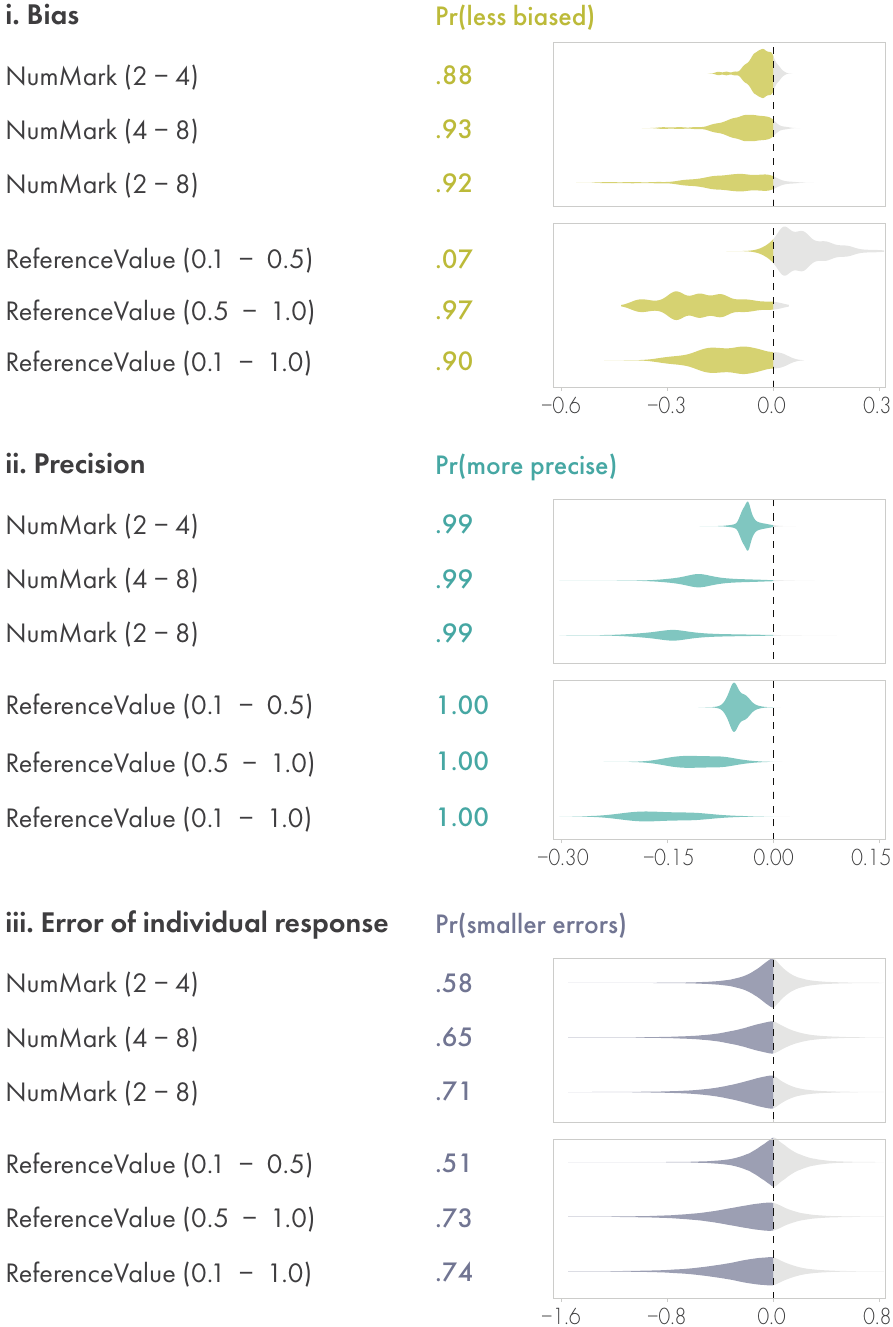}
\vspace{-18pt}
\caption{\textbf{The exmaples of quantified primary effects} of the number of marks and reference values. We take subtraction and calculate the marginal probabilities of being better (see Fig.~\ref{fig:primary-effects}b), averaging across visual channels and reference values (or different numbers of marks).}
\vspace{-12pt}
\label{fig:primary-effects-marginalized}
\end{figure}
 
\subsection{Primary effects}
 \label{sec:primary-effects}
The model suggests that the two experimental variables---the number of marks (\variable{NumMark}) and the reference value (\variable{ReferenceValue})---both have very strong effects on the reproduction responses across the three measures.
To show these effects, we take an average participant (to eliminate individual differences), conditional on the median trial (to rid learning/fatigue effects) and on the case where data mean is equal to the reference value (to remove the effects of the other marks in the same trial).   
 
 Fig.~\ref{fig:primary-effects}a shows all of the modeled effects, including the tendencies and the interactions between variables. Fig.~\ref{fig:primary-effects-marginalized} provides examples of quantified primary effects by showing how likely an average participant's responses are better (less biased/more precise/smaller errors).
 
\begin{minipage}[t]{10pt}
\begin{figure}[H]
\centering
\vspace{-13pt}
\hspace*{-25pt}\includegraphics[height=685pt]{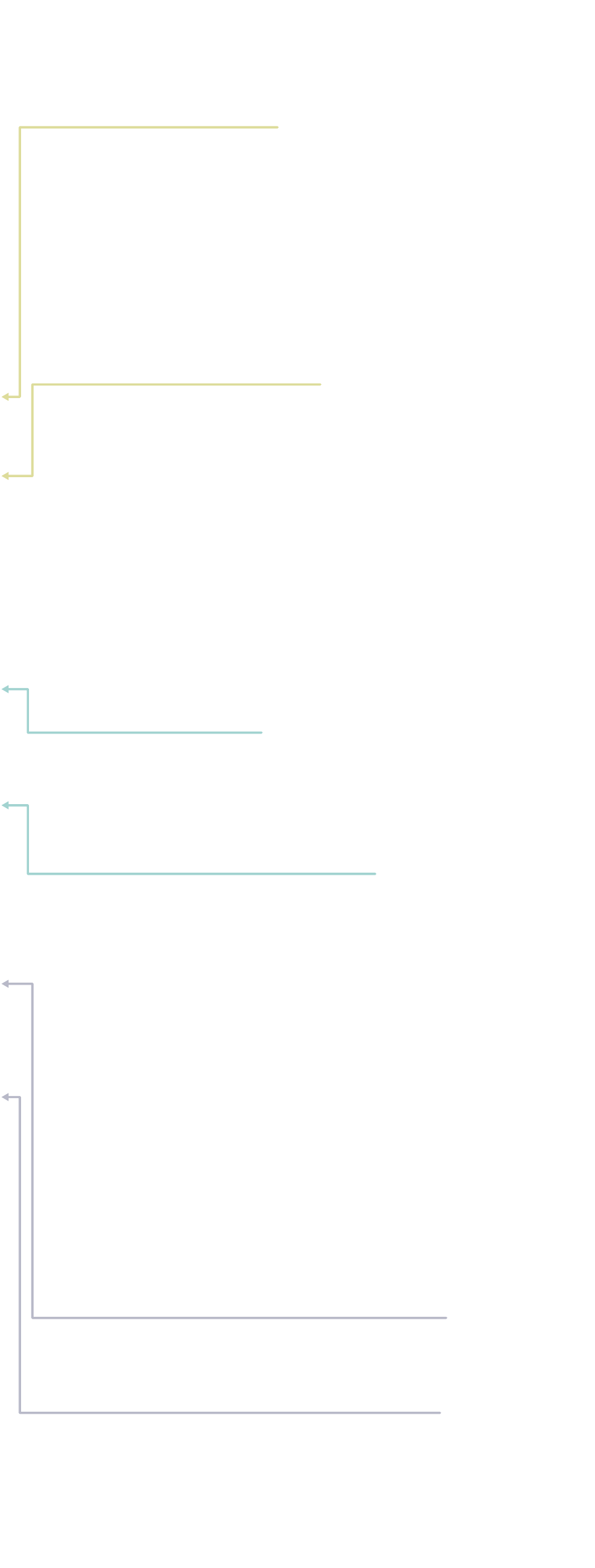}
\end{figure}
\end{minipage} 
\edef\myindent{\the\parindent}
\vspace{-5pt}
\hspace*{-16.5pt}\begin{minipage}[t]{250pt}\setlength{\parindent}{\myindent}

\paragraphhead{i. Bias} (Figs.~\ref{fig:primary-effects}i and~\ref{fig:primary-effects-marginalized}i) 
\addlessvspace

\noindent\mytag{Number of marks.} 
The average participant is very likely to be \textit{less} biased in the reproduction, when the number of marks is \textit{small}. 
For an average visual channel and an average reference value, the estimated probability that the average participant is less biased in a chart with 2 marks than with 8 marks is \biasFormattedText{.92}. 
That is, for the same reference value, we expect  92\% of responses with 2 marks to exhibit less bias than the responses with 8 marks. 

\addlessvspace

\noindent\mytag{Reference value.} 
The average participant is very likely to \textit{overestimate} a small reference value and seriously \textit{underestimate} a large reference value, and are least biased with a reference value around .4 or .5 (median).
For an average visual channel and an average number of marks, the estimated probability that the participant is less biased in a chart with the median reference value (.5) than the minimum value~(.1)  is .93 (this is 1-.07). 
Similarly, the estimated probability that an average participant is less biased in a chart with the median reference value (.5) than the maximum value (1.0) is \biasFormattedText{.97}.

\addlessvspace

\noindent\mytag{Interaction effects.}
The effects of \variable{NumMark} and \variable{ReferenceValue} interact, and each interacts with \variable{VisualChannel}.
For most of the visual channels but \condition{\mbox{position (line)}}, response bias increases when the number of marks is large and a reference value deviates from the median further.
Overall,  \condition{angle} is the visual channel where response bias is most sensitive to either a change in the number of marks or the  reference value; \condition{\mbox{position (line)}} is where bias is sensitive to the  reference value, but robust to the number of marks for large reference values.

\addmorevspace
\addmorevspace

\paragraphhead{ii. Precision}  (Figs.~\ref{fig:primary-effects}ii and~\ref{fig:primary-effects-marginalized}ii) 

\addlessvspace

\noindent\mytag{Number of marks.}
The average participant is very likely to be \textit{more} precise (more consistent) when the number of marks is \textit{small}.
For an average visual channel and an average reference value, the estimated probability that the participant is more precise with a chart of 2 marks than a chart of 8 marks is \precisionFormattedText{.99}. 

\addlessvspace

\noindent\mytag{Reference value.} 
The average participant is  \textit{more} precise with reproducing a  \textit{small} reference value and much less precise with reproducing a large reference value.
For an average visual channel and an average number of marks, the estimated probability that the participant is more precise with the minimum reference value (.1) than the median or maximum reference value (.5 or 1.0) is \precisionFormattedText{1.00} (nearly deterministic).

\addlessvspace

\noindent\mytag{Interaction effects.}
The effects of these two variables on precision interact with each other and further with visual channels.
Response precision is \textit{more} affected by the number of marks when the reference value is \textit{smaller}, except \condition{angle}, where precision is more affected by the number of marks when the reference value is \textit{large}.
Similarly, precision is \textit{more}  affected by the reference value with \textit{fewer} marks, except \condition{angle}, where precision is \textit{more}  affected by the reference value with \textit{more}  marks.
Overall, \condition{luminance} is the visual channel where precision is least sensitive to the reference value, and   \condition{\mbox{position (line)}} is where precision is most sensitive to the reference value.

\addmorevspace
\addmorevspace

\paragraphhead{iii. Error of individual response}  (Figs.~\ref{fig:primary-effects}iii and~\ref{fig:primary-effects-marginalized}iii) 
\addlessvspace

\noindent The samples drawn from the posterior distributions provide an estimation of errors in individual responses;  for the convenience of comparison, we took the absolute values.

\addlessvspace

\noindent\mytag{Number of marks.}
The average participant is likely to make \textit{smaller} errors with \textit{fewer} marks. 
For an average visual channel and an average reference value, the probability that a single future response exhibits a smaller error with 2 marks than with 8 marks is \responseFormattedText{.71}. 

\addlessvspace

\noindent\mytag{Reference value.} 
The average participant is likely to make smaller errors with a \textit{smaller} reference value.
The estimated probability that a single future response will have a  \textit{smaller}  error for the minimum reference value (.1) than the maximum (1.0) is \responseFormattedText{.74}. 

\addlessvspace

\noindent\mytag{Interaction effects.}
Reproduction error is likely affected by the number of marks slightly more in larger reference values for \condition{area} and \condition{angle}, less for \condition{\mbox{position (bar)}} and \condition{\mbox{position (line)}}, and similarly across different reference values for \condition{luminance} and  \condition{length}. 
These interactions effects are milder than those observed for bias and precision, owing in part to increased uncertainty in this measure relative to the aggregated properties described by bias and precision.
\end{minipage}

 \begin{figure*}[!b]
 \centering
 \vspace*{-12pt}
 \includegraphics[width=\textwidth]{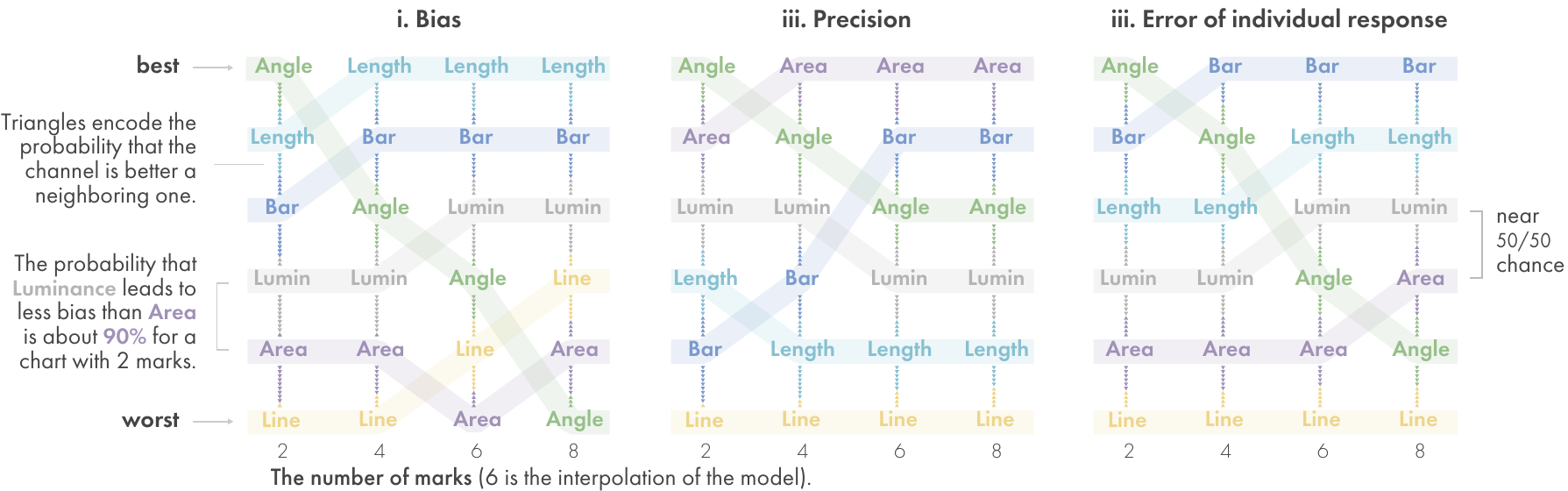}
 \vspace*{-20pt}
 \caption{\textbf{The probabilistic rankings of visual channels} on the three measures, augmented with the probability of the associated channel being better (less biased/more precise/smaller errors) than another.
 }
 \label{fig:ranks}
 \vspace*{-15pt}
\end{figure*}
\subsection{Secondary effects}
\label{sec:secondary-effects}
The model also suggests several moderate effects. To show the learning/fatigue effect, we condition on the average case where both reference value and the associated data mean are at their median (.5, .55, respectively). 
To show the effect of data properties (\eg the mean of all the data values in a trial), we condition on the average case where reference value is at its median (.5) in the median trial, and sampled all the possible values of data mean.
We also marginalize out the number of marks and visual channels and use an average participant.
\addmorevspace

\hspace*{-10pt}\begin{minipage}[t]{145pt}
\noindent\paragraphhead{i. Bias}
The participant appears to underestimate reference values at the beginning of the experiment.
In later trials, the participant generally increases the reproduced values and becomes less biased. 
In reproducing an average value, the participant seems not affected by other small reference values, but is likely to underestimate the median value when other reference values are large. 

\addmorevspace

\paragraphhead{ii. Precision}
The participant appears to become slightly less precise as the experiment goes on, possibly due to the fatigue effect.
In reproducing an average value, the average participant seems less precise when other larger values are present in the same trial; these larger values possibly distract the participant's judgment and reproduction.

\addmorevspace

\paragraphhead{iii. Error of individual response} 
It appears that learning or fatigue effects do not strongly affect response error.
In reproducing an average value, the participant is likely to make smaller errors when other reference values are small, and to make larger errors when other values are large.
The error of a response seems to largely increase when data mean is above .25, half of the median.
\end{minipage}
\hspace*{0pt}
\begin{minipage}[t]{80pt}
\begin{figure}[H]
\centering
\vspace{-21pt}
\includegraphics[height=300pt]{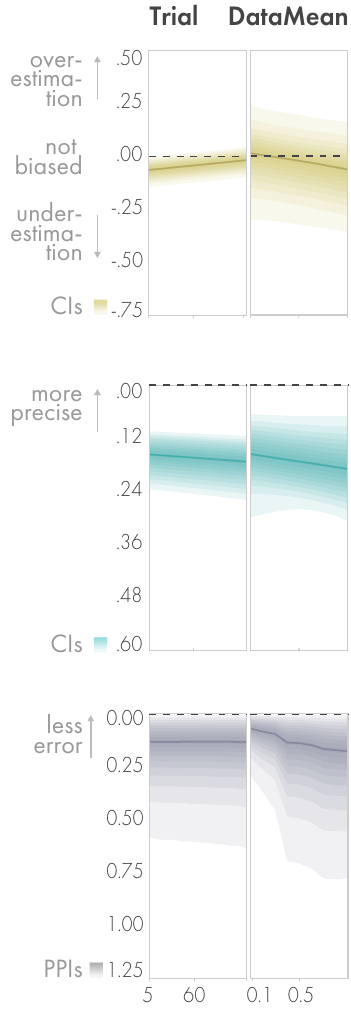}
\end{figure}
\end{minipage} 

\subsection{Deriving probabilistic ranks}
\label{sec:ranks}

\noindent One primary goal of this work is to derive ranks for the visual channels based on the reproduction task and compare them to those from two-value ratio judgment tasks~\cite{Cleveland1986,Heer2010,harrison2013influencing}.
A ranking list may provide a summary of effects for others to digest the results~(\eg \cite{harrison2014ranking,rosen2020linesmooth}).
However, a rank list may cause \textit{dichotomous thinking} (\eg ``A is always better than B.''), which belies the nuances in the ranking.
In the spirit of rethinking the previous ranks and in the context of Bayesian statistics, we will derive probabilistic ranks that acknowledge the \textit{uncertainty} in the observations and the modeling processes.


A rank in the probabilistic domain may mean that one is \textit{more likely} to be better than another. 
Hence we start by calculating these probabilities for visual channels. 
We marginalize all reference values and condition on the median trial with the median data mean.
We subtract the absolute values of each measure; 
if A is better (less biased/more precise/smaller errors) than B, we expect negative values after subtraction (see Fig.~\ref{fig:primary-effects}b). 
The estimated probability of A being better than B is the proportion of the negative values over the entire subtracted distribution.
When this probability is larger than 50\% (larger than chance or other thresholds), we say A is \textit{more likely} to be better than B, and A ``wins.''

We derive the probabilistic rank lists by pairwise subtraction and then build a chain of visual channels where any previous node on this chain \textit{always} ``wins'' any comparison to the later nodes for a given measure.
This is essentially a \textit{constraint satisfaction problem}~\cite{kumar1992algorithms,moritz2018formalizing}, and there are many methods to find a solution~\cite{kumar1992algorithms}. 
For the scale of our problem, we can build the chain by hand or apply heuristics (\eg sorting by how many times a visual channel ``wins'').
We construct a chain for each measure and visualize them in Fig.~\ref{fig:ranks}, augmented with the associated probabilities to convey uncertainty.

%


\textbf{The ranking produced by the precision of two-value ratio judgments does not hold} for each of the three measures nor any of the modeled numbers of marks.
The ranks changes with different numbers of marks across different measures, which suggests that the previous channel rank is likely not generalize to other visual comparison tasks.


\section{Discussion}
\label{sec:discussion}

The varying rankings and the effects of the number of marks and the reference values bring our discussion on the \textit{context} of a visualization below, which further invites a discussion on the implications of this work, along with an acknowledgment of the known limitations.  

\subsection{The context of a visualization}
\label{sec:context}

We find that showing more marks adds substantial noise to memory representations, and has an order of magnitude more influence on performance than the choice of channel. 
Squeezing more data into one visualization may cause viewers to increasingly  remember (and compare) data as statistics or rough global shapes \cite{brady2011hierarchical, chunharas2019adaptive}, rather than precise representations of individual values. As memory for each value likely also depends on its relation to the distribution of other values, as in work on neighborhood effects~\cite{zhao2019neighborhood,brady2015contextual, bae2017interactions} and distractor effects~\cite{Talbot2014FourEO}. 


We also indicate that the value of a mark (the reference value) has a more powerful effect on reproduction than the channel chosen: \textit{tall} bars are more biased than \textit{small} areas, even though \condition{position (bar)} is one of the least biased channels overall. 
The context of the reference value also indicates strong bias (\eg \condition{angle}), similar to past work where participants tend to be more biased and less precise with a value further from the ends of the range~\cite{mccoleman:2020:Bias, hollands2000bias} (\eg ``edge effects''); it also  aligns with psychophysical observations~\cite{gescheider2013psychophysics}, where low and high ends of the data range can serve as perceptual anchors (\eg ``the angle is 10\degree~from 90\degree~is perceptually congruent). 
Alternatively, for a channel like \condition{position (bar)}, participants perform better around the median value. This is probable that they resort to near-mean estimations when their memory falters, also consistent with better performance around the mean of the possible range of values~\cite{huang2019distinguishing}.

The mark-based reproduction task itself may also influence the context in the eye of the viewer. 
Line charts use the position channel but were always redrawn lower down than the reference values. 
They may be perceived as a single complex shape, or set of contrasting slopes, for the purposes of redrawing. Line charts indicate relative changes but may also create more bias in average value judgments for comparisons of lines with different baselines (consistent with~\cite{xiongEtAlAnchoring}). 
This suggests that the reproduction task may undervalue visual channels that provide good relative, but poor absolute information about the values.%

\begin{figure}[t!]
 \centering
 \includegraphics[width=.97\columnwidth]{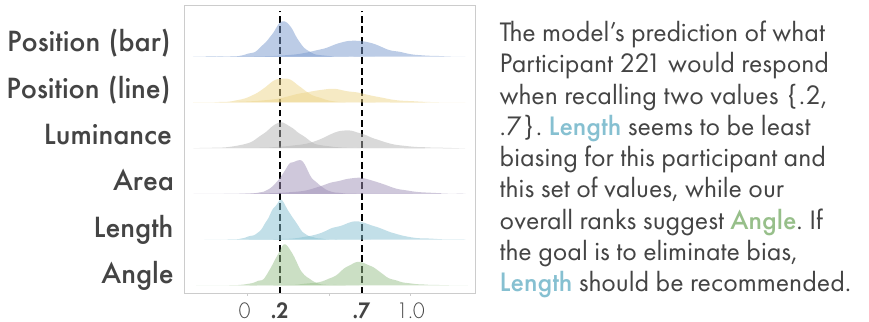}
 \vspace{-8pt}
\caption{An example of how the model could support design decision making via predicting a viewer's responses. 
} 
\vspace*{-13pt}
\label{fig:chooserApp}
\end{figure}

\subsection{Implications and future directions}
\label{sec:implications}

The previous channel ranking based on two-value ratio judgments appears to be attractive as a rule. That task \textit{feels} like a visual comparison distilled down to an atomic unit, which may lead to an assumption that a ranking based on that task should extrapolate to new ones. However, the present study denies this assumption (Fig.~\ref{fig:ranks}), with a different ranking for the two-value condition in a reproduction task. As such, \ul{designers ought to be increasingly skeptical of the channel ranking produced by two-value ratio tasks}~\cite{Cleveland1986}, which may not serve as a generalizable guideline for their usage. 

The present study also shows how the number of marks, the reference value, and other secondary factors such as data mean may strongly, progressively, and interactively affect reproduction.
Like previous work that had identified data category~\cite{Mackinlay1986} and distribution~\cite{kim2018assessing} as design factors, \ul{
other factors beyond visual channels could be critical inputs for design recommendations. 
}

Our rankings reveal the tendencies of the visual channels on different measures and the associated uncertainty, given the reproduction task.
If a designer is pursuing the exact or a similar task, these rankings could be used as a reference.
For example, \condition{length} and  \condition{\mbox{position (bar)}} generally lead to less bias and smaller errors, likely desired in reducing bias. 
\condition{Area} is surprisingly more precise but could lead to more bias and larger error,  likely preferred in improving precision. 
\condition{Luminance} is relevantly stable across different numbers of marks and measures, likely suitable when data size is varying.
\condition{Angle} seems sensitive to different numbers of marks and may be most useful for two marks.
\condition{\mbox{Position (line)}} seems ineffective in this reproduction task but may reduce bias for a larger dataset.
Knowing one's risk appetite for the misperception of \textit{bias} or \textit{precision} will inform the choice of visual channels.
Moving forward, \ul{parameterizing the influence of data properties (number of marks, values) and the designers' desire to optimize for lower bias, higher precision or lower error may help to inform visualization designers' decisions.}

Thinking of multiple factors may be difficult for designers, not to mention the possible conflicts and other specialized design considerations. 
Our analysis methods may shed light on resolving this complexity.
We were inspired by the recent modeling work~\cite{harrison2014ranking,kay2015beyond, szafir2017modeling,fernandes2018uncertainty, kale2020visual} and appealed to psychophysical laws~\cite{rensink2010perception,harrison2014ranking,yang2018correlation,mccoleman:2020:Bias}, entropy~\cite{rensink2016entropy,ryan2018glance,rosen2020linesmooth,chen2010information}, perceptual proxies~\cite{jardine2019perceptual,oberauer2013visual}, serial-position and ordering effects~\cite{guo2017shaping}, visual memory (\eg \cite{brady2015contextual, baddeley2001concept, miller1956magical}), neighborhood effects~\cite{zhao2019neighborhood}, and distractor effects~\cite{Talbot2014FourEO}. 
The final model used, incorporating empirical knowledge, is capable of providing preliminary recommendations given the inputs (see Fig.~\ref{fig:chooserApp}).
Thus, \ul{a modeling approach like ours may harmonize different factors and provide design candidates.}

\ul{It would be premature to derive other firm guidelines based on the present study}; additional studies will be needed or establish whether there exist broadly applicable guidelines. 
Our task relies on a purposely abstract reproduction task as a first step toward inspiring future work using more concrete comparison tasks.
Those studies will need to expand how the visualization research community operationalizes different visualizations tasks (\eg detecting trends and motifs, immediate or later comparison,  and viewing a visualization with thousands of data points),  and what  `good performance' means in a task (\eg precision, bias, error, speed~\cite{saket2018task,kim2018assessing}, etc.).
In addition, other factors such as top-down effects like prior knowledge~\cite{xiong2020curse} and expectations~\cite{peck2013using,kim2017gap} may impact reproduction task performance, and individual differences~\cite{ottley2015personality} and spatial ability~\cite{ottley2016bayesian} may affect strategies that subsequently impact task performance, which could be promising directions.

\begin{figure}[t!]
\centering
\vspace{-2pt}
\includegraphics[width=\columnwidth]{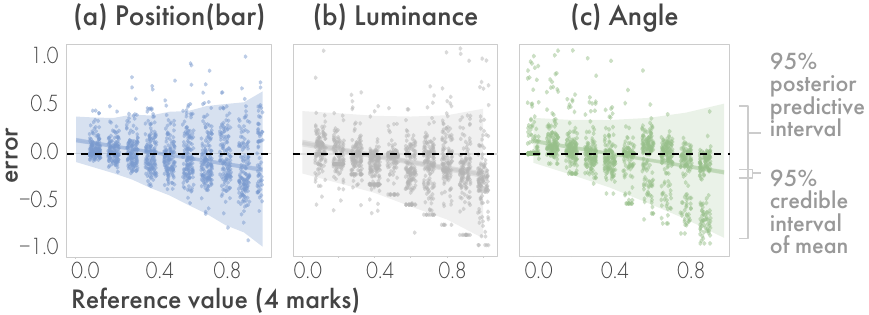}
\vspace{-17pt}
\caption{The moderate non-linearity and symmetry in responses for (b) \condition{luminance} and (c) \condition{angle}, compared to (a) \condition{\mbox{position (bar)}}. 
}
\vspace{-12.5pt}
\label{fig:circ-effects}
\end{figure}

\subsection{Limitations}
\label{sec:limitations}
For the sake of comparability, we treated all visual channels equally in designing the experiment and analyzing the data. This makes the channels easier to compare, potentially at the cost of the usability for some. 
The redrawing method may add noise and bias to data, as it might not be equally intuitive for all the visual channels. For example, for  \condition{angle}, we mapped the $y$-coordinate of the cursor to the degrees of the angle, which might be more difficult to draw than others (\eg \condition{\mbox{position (bar)}} that maps the cursor to the height of the bar). 
Similarly, always dragging up from the zero value might result in a bias towards smaller values for the two position channels, possibly explaining the underestimation in \condition{position (line)} noted above. These response methods likely have an impact on the result, and a comparison of different response methods will be critical to generalization of these results.
The model also always assumes linearity between errors of the responses and all the variables.   
While most of the data meet this assumption, visual channels like \condition{angle} display moderate non-linearity across different reference values (Fig.~\ref{fig:circ-effects}), likely affecting the estimation of the model.

\section{Conclusion}
\label{sec:conclusion}

We revisited the ranking of visual channels~\cite{Cleveland1984} using a visual reproduction task  as a proxy of various visual comparison tasks.
We tested participants' reproduction performance with six visual channels: \condition{\mbox{position (bar)}}, \condition{\mbox{position (line)}}, \condition{luminance}, \condition{area}, \condition{length}, and \condition{angle} across different numbers of marks and data values.
With a Bayesian multilevel model, we show that both the number of marks and the reference value strongly affect the bias and precision in a set of responses, as well as errors of individual responses; the number of marks gradually dominates the differences in visual channels and reference values, reflecting a strong limit on working memory, that likely serves to limit most comparison tasks in data visualization. 
We further derive probabilistic rankings from the model for each measure and show that the previous ranking \cite{Cleveland1986} does not hold. 
We demonstrate the limitations of the previous ranking~\cite{Cleveland1984}, offer the preliminary new rankings based on a reproduction task, and present a Bayesian modeling approach to rank visual channels, all for future work to continue exploring this area.


\acknowledgments{
Thank you to Satoru Suzuki and members of the Visual Thinking Laboratory at Northwestern University for their suggestions during the experimental design. The authors also thank the anonymous reviewers for their feedback. This work was supported in part by grants BCS-1653457 and IIS-1901485 from the National Science Foundation.
}

\bibliographystyle{abbrvdoi}
\bibliography{setSize-bib.bib}

\cleardoublepage

\appendix
\newpage
\pagenumbering{arabic}

\onecolumn
\setcounter{page}{1}
\setcounter{section}{0}
    
\counterwithin{figure}{section}
\counterwithin{table}{section}

\begin{center}
\huge{\textsf{\thetitle}}

\vspace*{10pt}

\Large\textsf{Appendices}

\vspace*{10pt}

\end{center}

\section{Visual Channels Details}
\label{sec:app-channel-details}
In all the conditions, for all the \variable{visual channels}, participants redrew the stimulus by dragging their mouse above the data mark. 
This may be obvious for a visual channel such as a bar graph, where the height of the bar reflects the underlying value. 
It is a little less obvious for a wind map, where the orientation of the line reflects the underlying value. 
The motivation for this consistent response is to ensure that differences in the observed errors are not because of different motor demands. Using a consistent response method across different visual channels may introduce variance in the data.

For all visual channel types, the data marks were presented within an axis that was nearly half the height of a 23'' screen. Running with a resolution of 1280 $\times$ 800 pixels, the usable Y range for each data display was $\frac{800}{2}$ - 20 pixels. The 20 pixels was dynamically determined ($\frac{1}{40}$ of the height) to keep the highest values from ever hitting the top of the screen. Similarly, the $x$-axis was determined to be half of the width, minus an edge buffer ($\frac{1}{30}$ of the width) to keep the data marks from hitting the edge of the screen. The maximum height for a bar is 390 pixels. The bars were 1/16th of the $x$-axis wide (37.3 pixels). As with all of the data marks used in the current experiments, participants were presented with a random selection of values from 0.1 - 1.0. The minimum bar height (0.1) was then 39 pixels.

\addmorevspace

\paragraphhead{Experiment A: The visual channels with a common baseline}

The bottom of the each visual mark in the first experiment was randomly selected within the y axis range for each trial, so participants could not rely solely on position to remember the values they were shown in the graph. 
The background for all conditions was light grey (25\% black).

\addmorevspace

\paragraphhead{Position-bar}%
Participants respond using a computer mouse and clicking/dragging above the data mark to draw it to the size they remember seeing in the initial data presentation. Note that the initial value was 0 for the bar. For this, and all conditions, the response was initialized to the lowest value in the range of possible responses. Participants can click or adjust the same mark multiple times. Their response is fully self-timed. If participants click away from the indicated response space, the graph briefly flashes off the screen to provide unintrusive feedback about the viable response regions. 

\addmorevspace
\paragraphhead{Position-line}%
The heights of the points on the line were the same as the heights of the bars in the position condition, such that the maximum height for a line chart vertex was 390 pixels (the height of the Y range).  The stimulus was created by joining the randomly selected point heights. The line was four pixels wide. 

Participants respond by re-drawing the line in the same manner as the bars. They click or drag the point above the zero-mark to adjust the height of the line, in an attempt to match the line chart they saw in the initial stimulus as closely as possible.

\addmorevspace
\paragraphhead{Luminance}%
The heat map marks were 37 $\times$ 37 pixels (same width as the bars). While the aligned bars and the line chart represented changing values by changing in height, the heat maps changed how light/dark the presented marks were. The maximum value was white, such that as the participant dragged their mouse higher, the box that they were adjusting became lighter. 0\% was represented the same color as the stimuli in the other conditions: RGB values [127.5, 127.5, 127.5] or 50\% black. 

Participants respond by click-and-dragging above the data point just as in the bar and line graphs. That is, to make the mark darker, they drag the mouse cursor higher on the screen. Just as with the bar and line charts, the initial response started at 0, and participants had to drag the mouse to change the value of the data mark. The only difference between the heat map and the previous conditions' response method, is that, for the heat maps, when participants dragged the mouse up and down, the position of the mark stayed constant and the color changed; in the line and bar charts, when the participants dragged the mouse up and down the position of the mark changed and the color stayed constant.

\addmorevspace

\paragraphhead{Experiment B: The misaligned charts}

\addmorevspace
\paragraphhead{Length}
As with the other visual channels in this experiment, the misaligned bars do not share a common baseline. Since the baseline varies within the $y$-axis range, the maximum height of these bars is reduced so that there is enough room to have the 0.1 - 1.0 variation while still having the participants respond in the opposite corner from where the bars were originally presented. Otherwise, the bars were the same as in Experiment 1A: the aligned bars condition. 

\addmorevspace

\paragraphhead{Angle}
The wind map line segments were as long as the bars were wide ($\frac{1}{16}$ of the $x$-axis). The maximum value (1.0) was presented by a horizontal line (180\degree). The remaining presented values (0.1 - 1.0) were proportions of the 0-180\degree range, such that 0.1 was 18\degree. 

Participants make their responses just as in the other conditions: by adjusting the height of the mouse cursor above the data mark. Because there is 180\degree of possible response space it could be difficult to tell whether a nearly-flat was close to 0 or 180\degree. To address this concern for participants, we ensured that they were well trained before they began the task, and that there was an origin for the line segment, such that a line falling to the right was closer to 0\degree and a line falling to the left was closer to 180\degree. Just as with the other visual channels, participants' response screens were initialized to 0, so it was clear to participants that the initial screen was 0 degrees, they would exceed the 0.5 response only adjusting the line past 90\degree.

\addmorevspace

\paragraphhead{Area}
The area charts used circle visual channel, where the area of the circle mark changed in direct linear proportion to the size of the presented value. The maximum value (1.0) was represented by a circle with a diameter the same width as the bars ($\frac{1}{16}$ of the $x$-axis, or 37.3 pixels). The remaining values were represented as a proportion of that circle's area, such that the radius for the circle representing 0.1 was $\frac{\sqrt{0.1 * maximum\:area}}{\pi}$.

Participants re-drew the area values just as they re-drew the bar, line, and heatmap above: they dragged their mouse on top of the data mark to make their response. Note that the adjustment of this mark was scaled to the change the area of the circle (not, the diameter) since people tend to perceive the differences in the area of the circle as the natural data mapping. One unit increase in mouse cursor height, then, corresponds to one unit increase in mark area.

\section{Alternative Modeling Approaches}
\label{sec:div-by-presented-values}
\begin{figure*}[tbp!]
 \centering
 \includegraphics[width=.825\textwidth]{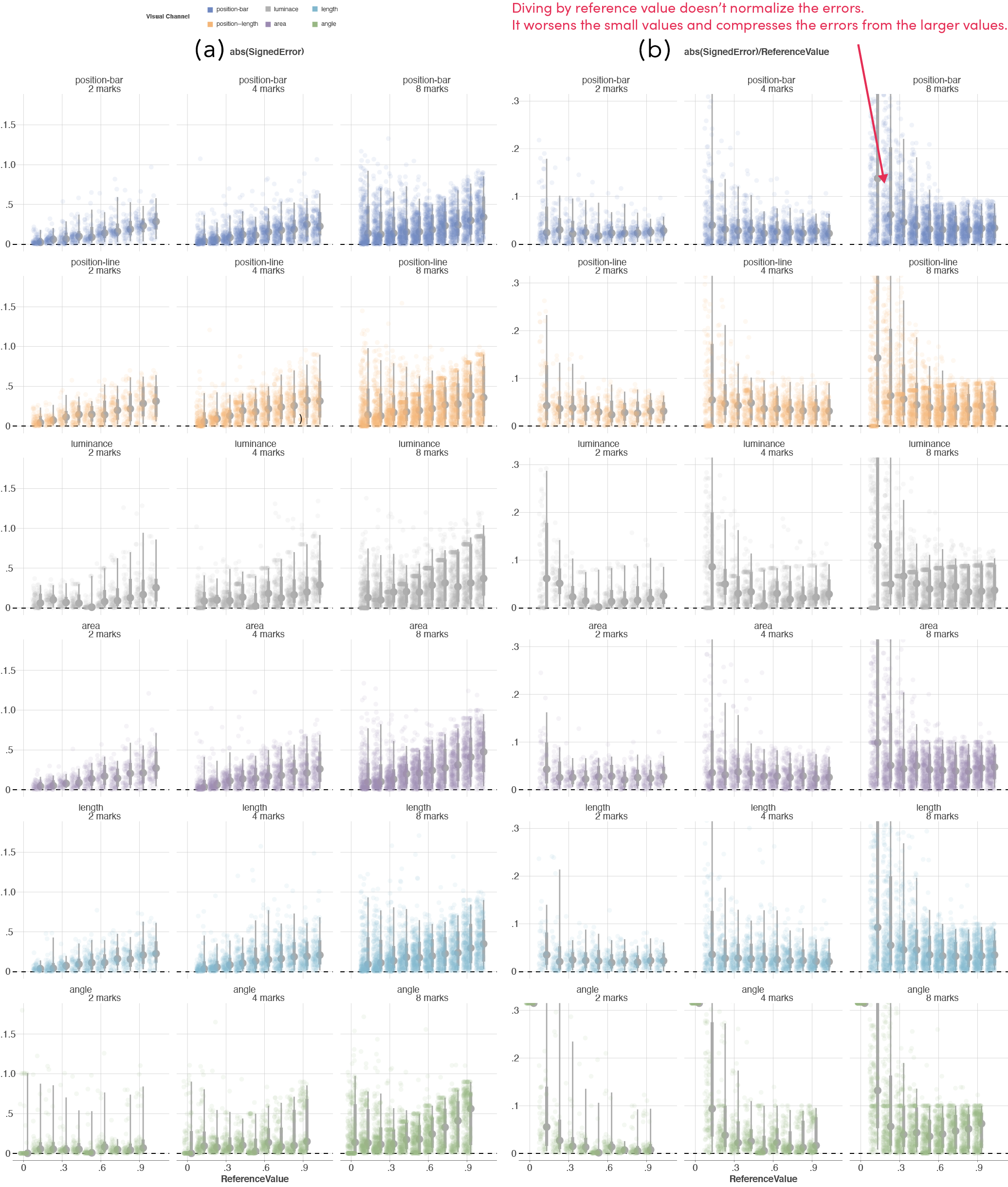}
  \vspace{-10pt}
 \caption{Alternative approaches of  using our knowledge about perception. We could transfer data by dividing reference value, but this still did not normalize error distributions. Bayesian approaches can directly model skewed distributions. So, using reference value as a predictor is rational.}
 \label{fig:alternative-approaches-adding-presentedvalue}
 \vspace{-10pt}
\end{figure*}

An alternative approach to incorporate \variable{ReferenceValue} as a predictor is to divide all the responses by their corresponding presented values. 
This approach assumes that each visual channel follows Weber's law, and therefore division is able to normalize errors. 
However, this assumption is too strong, and we found that after dividing presented value, errors still vary with presented value, and errors for small presented values were exacerbated.
Fig.~\ref{fig:alternative-approaches-adding-presentedvalue}a shows absolute errors.
Fig.~\ref{fig:alternative-approaches-adding-presentedvalue}b shows absolute errors divided by presented value, where errors are still non-linear.


\section{Calculating LogAbsError (Ratio) for Our Data}

%

We replicate Cleveland and McGill's analysis to facilitate comparison.
Cleveland and McGill's study and their successors based the ranks of visual visual channels on a task of ratio estimation and the log-transformed absolute errors.
\begin{align*}
    & LogAbsError = log_2(| bias\ of\ percentage\ error | \  + \ .125)
\end{align*}%

We utilize the same ratio measure to  the modeled bias  in our experiment and its reference:
\begin{align*}
    & LogAbsError= log_2(\bigg| \dfrac{bias}{Reference}\bigg| * 100 \  + \ .125)
\end{align*}%
We apply this measure to our modeled bias  and calculate $t$ confidence intervals. 
Since the model has 49 participants levels, we can consider that we have 49 participants.
The results are used to generate Fig.~\ref{fig:teaser}.

\end{document}